%% file: prb.tex
\documentclass[english,prb,aps,twocolumn, 10pt]{revtex4-2}

\usepackage{amsmath}
\usepackage{amssymb}
\usepackage{graphicx}
\usepackage[english]{babel}
\usepackage{array}
\usepackage{verbatim}
\usepackage[colorlinks=true, pdfstartview=FitV, linkcolor=blue, citecolor=blue, urlcolor=blue]{hyperref} % enable links
\usepackage{oplotsymbl} % special symbols, like \pentago

\usepackage{multirow} % for Table 1
\usepackage{siunitx} % collides with our command \unit, so renamed it below.
\usepackage{physics}
\usepackage{chemformula} % used for \ch{NV-}, \ch{NV^0} only

\def\second   {{2$^{\rm nd}$ \/}}

\def\etal     {{\it et al.}}

\newcommand{\unitformat}[1]{\,\mathrm{#1}}

\newcommand{\kB}{k_{\rm B}}
\newcommand{\muB}{\mu_{\rm B}}

\newcommand{\degree}{^\circ}

\newcommand{\eII}{NV-2}

\newcommand*\tcircle[1]{%
 {\raisebox{-0.25pt}{%
    \textcircled{\scriptsize #1}}
  }%
}

\newcommand{\dperp}{\delta_\perp}
\newcommand{\Ex}{\text{E}_\text{x}}
\newcommand{\Ey}{\text{E}_\text{y}}
\newcommand{\ms}[1]{$m_{\mathrm{S}} = #1$}
\newcommand{\PLms}[1]{$\text{PL}_{m_{\mathrm{S}} = #1}$}
\newcommand{\x}{\text{x}}
\newcommand{\y}{\text{y}}
\newcommand{\mS}{m_{\mathrm{S}}}

\newcommand{\kcps}{\text{kcps}}

\DeclareSIUnit{\rad}{rad}

\begin{document}
\title{Modeling temperature-dependent population dynamics in the excited state of the nitrogen-vacancy center in diamond}

\author{S.~Ernst$^{1,\dagger}$, P.~J.~Scheidegger$^{1,\dagger}$, S.~Diesch$^{1}$, and C.~L.~Degen$^{1,2}$}
\email{degenc@ethz.ch}
\thanks{$^\dagger$These authors contributed equally.}
\affiliation{$^1$Department of Physics, ETH Zurich, Otto Stern Weg 1, 8093 Zurich, Switzerland.}
\affiliation{$^2$Quantum Center, ETH Zurich, 8093 Zurich, Switzerland.}
\date{\today}
	
\begin{abstract}
The nitrogen-vacancy (NV) center in diamond is well known in quantum metrology and quantum information for its favorable spin and optical properties, which span a wide temperature range from near zero to over 600 K. Despite its prominence, the NV center's photo-physics is incompletely understood, especially at intermediate temperatures between 10-100 K where phonons become activated.
In this work, we present a rate model able to describe the cross-over from the low-temperature to the high-temperature regime. Key to the model is a phonon-driven hopping between the two orbital branches in the excited state (ES), which accelerates spin relaxation via an interplay with the ES spin precession. We extend our model to include magnetic and electric fields as well as crystal strain, allowing us to simulate the population dynamics over a wide range of experimental conditions.
Our model recovers existing descriptions for the low- and high-temperature limits and successfully explains various sets of literature data. Further, the model allows us to predict experimental observables, in particular the photoluminescence (PL) emission rate, spin contrast, and spin initialization fidelity relevant for quantum applications. Lastly, our model allows probing the electron-phonon interaction of the NV center and reveals a gap between the current understanding and recent experimental findings.
\end{abstract}

\maketitle

%%%%%%%%%%%%%%%%%%%%%%%
\section{Introduction}
\label{sec:intro}
The negatively charged nitrogen-vacancy (\ch{NV-}) center in diamond has become one of the most studied solid-state spin defects~\cite{doherty13} owing to its wide range of potential applications in various quantum technologies~\cite{childress13,schirhagl14}.  Essential for its success is an extraordinarily long spin coherence time~\cite{herbschleb2019} in the ground electronic state (GS) combined with an optical preparation and readout of the spin state~\cite{gruber97, jelezko04observation, shields15, hopper16, robledo11nature, irber21}, which are maintained even at  ultra-low~\cite{scheidegger22} and well above room temperature~\cite{toyli12}.
The optical preparation and readout of the spin state relies on an optical cycle that is mostly spin conserving~\cite{robledo11njp, manson06}.  Upon optical excitation, the excited electronic state (ES) decays either radiatively by emission of a photon, or non-radiatively via an inter-system crossing (ISC) to a metastable singlet state~\cite{doherty13}.  Since the ISC is strongly spin-dependent, the \ch{NV-} spin states have different average emission rates and prolonged illumination leads to preferential population of the \ms{0} spin state.

Although this basic mechanism for spin contrast and spin polarization is well known, the detailed population dynamics in the ES are not entirely understood~\cite{gali19}.  In particular, transitions between the ES sublevels show a marked temperature dependence that can lead to rapid spin relaxation and  impair the optical readout.  At cryogenic temperature, the ES is known to be an orbital-doublet, spin-triplet consisting of six distinct sublevels~\cite{tamarat08, batalov09, manson06}.   At room temperature, the ES is averaged to an effective orbital-singlet, spin-triplet~\cite{doherty13} through phonon-driven transitions~\cite{rogers09, batalov09, plakhotnik14, fuchs10}.  This phonon-interaction within the ES has been attributed to a dynamic Jahn-Teller effect~\cite{fu09, abtew11}.  The phonon-induced mixing rate of the vibronic states of the ES has been determined in cryogenic experiments~\cite{goldman15, goldman15prb, goldman_erratum_2017}.  However, the lack of a fundamental model unifying the rate models at cryogenic~\cite{happacher22, rogers09, goldman15prb} with those at room temperature~\cite{tetienne12, robledo11njp} remains a central gap in the understanding of NV center population dynamics.

In this work, we present a comprehensive model of the \ch{NV-} center's spin dynamics spanning the full temperature range from the low to the high temperature regime.  We show that the interplay of spin and orbital dynamics in the ES leads to rapid spin relaxation at intermediate temperatures, even for NV centers with zero intrinsic strain.  We further include the effects of applied magnetic and electric bias fields, as well as crystallographic strain, on the level structure and population dynamics.
These parameters constitute key quantities in metrology~\cite{maze08, balasubramanian08, dolde11, kucsko13, acosta10} and have previously been used to fine-tune the \ch{NV-} center properties~\cite{bassett11, lee16} for information technology.
Using a master equation approach, we simulate key experimental observables, including the photoluminescence (PL) emission rate, the spin contrast, and the spin-state readout fidelity relevant for quantum applications.
Finally, we introduce a measurement approach to probe electron-phonon interactions and contributing phonon modes, allowing us to link recent experimental findings~\cite{rogers09, ernst2023, blakley2023_arxiv, happacher2023} with previous theoretical studies~\cite{abtew11, plakhotnik15, goldman15, goldman15prb, goldman_erratum_2017}.

Our paper is structured as follows:
In Sec.~\ref{sec:model} we present our numerical model.
We discuss the level structure at low temperature (Sec.~\ref{sec:H}),
introduce classical transition rates to model the population dynamics under optical excitation (Sec.~\ref{sec:rates}), 
and add phonon-induced transition rates between orbital states to include the influence of temperature (Sec.~\ref{sec:phonon_rates}).
We then implement this rate model using the master equation approach (Sec.~\ref{sec:ME}), and describe how to obtain experimental observables (Sec.~\ref{sec:SNR}).
In Sec.~\ref{sec:sim} we use our model to simulate the temperature dependence of the ES spin dynamics. We identify a prominent spin relaxation process at intermediate temperatures (Sec.~\ref{sec:temp}), which we verify with independent Monte Carlo simulations. We then discuss implications of the spin relaxation process for spin-state initialization and readout (Sec.~\ref{sec:implications_SNR}), and explain dependencies on magnetic field (Sec.~\ref{sec:B}) and crystal strain or electric field (Sec.~\ref{sec:strain}).
Finally, in Sec.~\ref{sec:experimental}, we put our model in context with previous work.
We verify the model over a wide temperature range (Sec.~\ref{sec:experimental_pl}), from low strain (Sec.~\ref{sec:comparisonISC}) to very high strain (Sec.~\ref{sec:trf}),
and reveal a gap in the current understanding of the electron-phonon coupling (Sec.~\ref{sec:probing}).
We summarize our findings and implications for future work in Sec.~\ref{sec:conclusion}.

%%%%%%%%%%%%%%%%%%%%%%%
\section{Rate Model}
\label{sec:model}

\subsection{Level structure}
\label{sec:H}
\begin{figure}
    \includegraphics{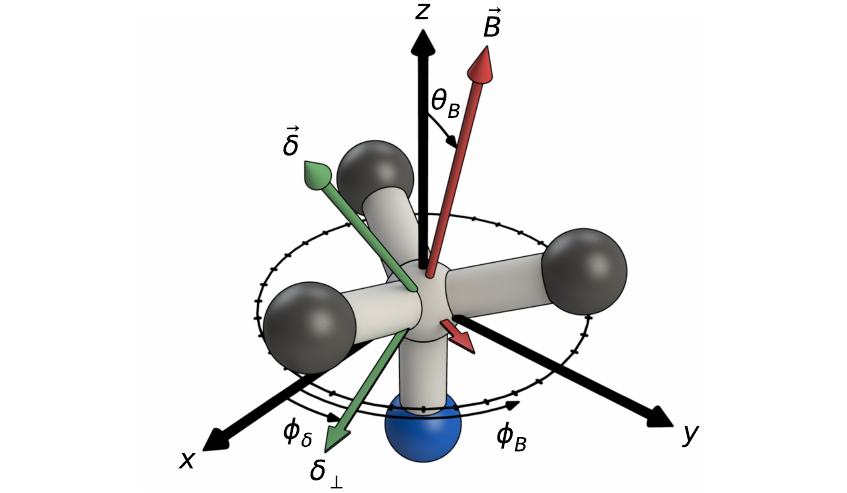}
	\caption{NV center coordinate system used in this work. Carbon atoms are shown in black, the substitutional nitrogen atom in blue, and the vacancy in gray. The $x$-axis is along one of the carbon bonds.
    $\vec{B}$ is the applied magnetic field (red), $\vec{\delta}$ the strain or electric field (green), and $\delta_\perp$ its in-plane component. The remaining symbols are explained in the text.
	} 
	\label{fig:0}
\end{figure}
We begin by introducing the coordinate system of the NV center in the diamond lattice in Fig.~\ref{fig:0}.  
We define the $z$-axis by the unit vector along the symmetry axis of the NV center pointing from the nitrogen atom towards the vacancy site, and the $x$-axis by one of the three in-plane vectors pointing from the vacancy to one of the carbon atoms~\cite{doherty13,huxter23}.
The in-plane angles of the magnetic field $\phi_B$ and the strain field $\phi_\delta$ are defined relative to this $x$-axis in the $xy$-plane.
An electric field is equivalent to a strain field~\cite{doherty13}. The magnetic field misalignment angle $\theta_B$ is measured with respect to the $z$-axis.

\begin{table}[t]
	\centering
	\begin{tabular}{l l l}
		\multicolumn{3}{l}{\textbf{Interactions}} \\ [0.5ex]
		$ D_{\text{GS}} $ & GS spin-spin int. & $ \SI{2.878}{\giga\hertz} $~\cite{chen11apl} \\
		\hline
		$ g_{\text{GS/ES}} $ & GS/ES el. $ g $-factor & $ \num{2.003} $~\cite{doherty13} \\
		\hline
		$ D_{\text{ES}}^\parallel $ & ES axial spin-spin int. & $ \SI{1.44}{\giga\hertz} $~\cite{bassett14} \\
		\hline
		$ D_{\text{ES}}^\perp $ & ES trans. spin-spin int. & $ \SI{1.541}{\giga\hertz}/2 $~\cite{bassett14} \\
		\hline
		$ \lambda_{\text{ES}}^\parallel $ & ES axial spin-orbit int. & $ \SI{5.33}{\giga\hertz} $~\cite{bassett14} \\
		\hline
		$ \lambda_{\text{ES}}^\perp $ & ES trans. spin-spin/orbit int. & $ \SI{0.154}{\giga\hertz} $~\cite{bassett14} \\
		\hline
		$ g_{l} $ & ES orb. $ g $-factor & $ \num{0.1} $~\cite{rogers09} \\ [1.5ex]
		
		\multicolumn{3}{l}{\textbf{Fields}} \\ [0.5ex]
		$ \dperp $ & ES in-plane strain/el. field &  $ \SI{40}{\giga\hertz} $ (default)  \\
		\hline		
		$ \phi_\delta $ & strain/el. field in-plane angle &  $\SI{0}{\degree}$ (default) \\
		\hline
		$ B $ & mag. field magnitude &  $   \SI{0}{\tesla} $ (default)  \\
		\hline
		$ \theta_B $ & mag. field misalignment angle &   $ \SI{0}{\degree} $ (default)  \\
		\hline
		$ \phi_B $ & mag. field in-plane angle &  $\SI{0}{\degree} $ (default) \\ [1.5ex]
		
		\multicolumn{3}{l}{\textbf{Rates}} \\ [0.5ex]
		$ k_r $ & opt. emission rate &  $ \SI{55.7}{\per\micro\second} $ \\
		\hline
		$ k_{E_{12}} $ & avg. ISC rate for $ \mS=\pm 1 $ &  $ \SI{98.7}{\per\micro\second} $ \\
		\hline
		$ k_{E_{\text{xy}}} $ & ISC rate for $ \mS=0 $ & $ \SI{8.2}{\per\micro\second} $ \\
		\hline
		$ r_\beta = $ &$  \beta_\x/\beta_\y $ opt. excit. branching ratio & $ \num{1} $ \\
		\hline
		$ r_S = $ &$ k_{S0}/k_{S1}$ SS branching ratio & $ \num{2.26} $ \\
		\hline
		$ \Delta E $ & SS emitted phonon energy & $ \SI{16.6}{\milli\electronvolt} $~\cite{robledo11njp} \\
		\hline
		$ \tau_{S,0} $ & SS decay time at $ T = \SI{0}{\kelvin}$ & $ \SI{320}{\nano\second} $ \\
		\hline
		$ T $ & temperature &  $ \SI{0}{\kelvin} $ (default) \\
		\hline
		$ \eta $ & ES el.-phonon coup. strength & $ \SI{176}{\per\micro\second\per\cubic\milli\electronvolt} $ \\
		\hline
		$ \Omega $ & phonon cutoff energy & $ \SI{168}{\milli\electronvolt} $ \\  [1.5ex]
		
		\multicolumn{3}{l}{\textbf{Setup}} \\ [0.5ex]
		$ P $ & laser power & $ \SI{2.34}{\milli\watt}$ (default) \\
		\hline
		$ b $ & background & $ \num{27.5}\,\kcps\,\si{\per\milli\watt} $ \\
		\hline
		$ R $ & collection over excit. eff. & $ \num{88.4}\,\kcps\,\si{\milli\watt\micro\second} $ \\
		\hline
		$ A $ & opt. alignment (excit. eff.) & $ \SI{136}{\per\watt} $ \\
	\end{tabular}
	\caption{
            Overview of the parameters used in our rate model. The given values are based on fit results of sample \eII{} in a nanostructured pillar in  Ref.~\cite{ernst2023} and additional literature. For parameters varied throughout this work, default values are specified.}
	\label{tab:1}
\end{table}

\begin{figure*}
    \includegraphics{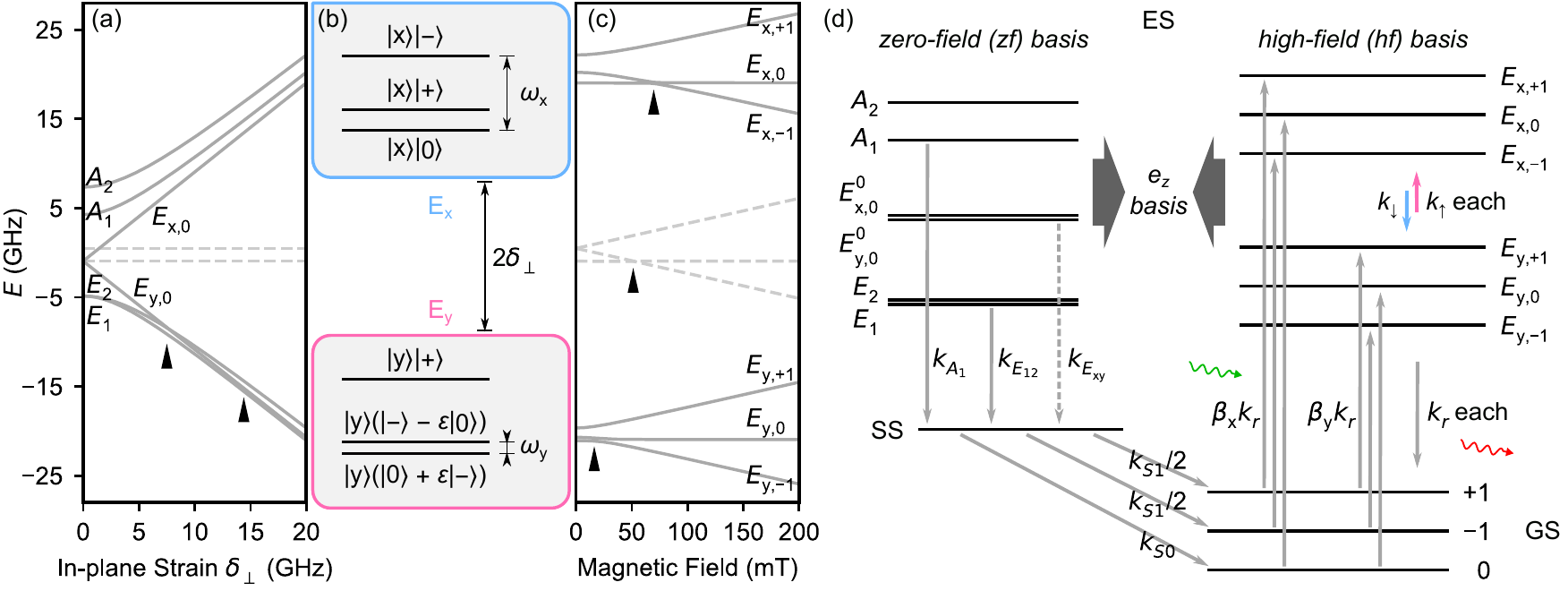}
	\caption{(a) Energy levels of the ES as a function of strain at zero magnetic field. In (a) and (c), level names are printed where they are approximately eigenstates, and level-anti-crossings (LAC) are indicated by black markers. The dashed lines are the eigenlevels of $\mathcal{H}_{\text{spin}}$, obtained by a partial trace over the orbital sub-space. They constitute the observed ES energy levels at room temperature.
    (b) Under strain, two orbital branches $\Ex$ and $\Ey$ form, split by approximately $2\delta_\perp$. The eigenstates of $ \mathcal{H}_{\text{orbit}} \otimes \mathcal{H}_{\text{spin}}$ are composite states of the orbital states $\ket{\x}$ and $\ket{\y}$ and superpositions of the spin states $\ket{0}$ and $\ket{\pm} \propto (\ket{+1}\pm\ket{-1})$. Only two spin states mix significantly in this example, with $\varepsilon = \epsilon_{\ket{\y}\ket{0}, \ket{\y}\ket{-}}$. Level spacings $\omega_{\x/\y}$ are given for the discussion in Fig.~\ref{fig:2}.
    (c) Energy levels of the ES as a function of axial magnetic field at $\delta_\perp = \SI{20}{\giga\hertz}$. Under high strain and magnetic field, eigenstates of orbit and spin are formed, for example $\ket{E_{\x,+1}} = \ket{\x}\ket{+1} $.
    (d) Rate model employed in this work.
    The rates listed in Tab.~\ref{tab:1} and the text are depicted by arrows.
    The dashed arrow indicates a low decay probability, allowing for the optical initialization and readout of the GS spin state. Excitation by green light and emission of red light are indicated by wavy arrows.
    The two bases \textit{zf} (see left in (a)) and \textit{hf} (right in (c)) are connected via basis transformations (dark gray arrows) to the $e_z$ \textit{basis} of the ES Hamiltonian.
	} 
	\label{fig:1}
\end{figure*}

Next, we introduce the Hamiltonian of the \ch{NV-} center.
The detailed level structure is discussed in Ref.~\cite{doherty13}.
The Hamiltonian of the $^3\text{A}_2$ electronic GS is a spin-triplet-orbital-singlet, and is given by
\begin{equation}\label{equ:HgsNoStrain}
		\hat{H}_{\text{GS}}/h =
		D_{\text{GS}} \left( \hat{S}_z^2 - \tfrac{2}{3} \, \hat{\mathbb{I}}_3 \right)
		+ \muB g_{\text{GS}} \hat{\vec{S}} \cdot \vec{B}\,,
\end{equation}
where we neglect hyperfine interactions and electric and strain fields, since their influence on the GS spin states is minor.
Here, $\hat{S}_i $ are spin--1 operators and $\hat{\mathbb{I}}_3 $ is the associated identity operator, $\vec{B}$ is the external magnetic field, $\muB$ is the Bohr magneton, and $h$ is the Planck constant which we use in units of $\si{\joule\per\hertz}$.  Values for the GS zero-field splitting $D_{\text{GS}}$ and the $g$-factor are listed in Tab.~\ref{tab:1}. 
Note that $D_{\text{GS}}$ has a slight temperature dependence~\cite{acosta10, lourette23}
which we neglect here as it has no influence on the population dynamics.  Equally, we neglect the mild temperature and strain dependence~\cite{plakhotnik14} of all other interactions listed in Tab.~\ref{tab:1}.

The Hamiltonian of the $^3\text{E}$ electronic ES is a spin-triplet-orbital-doublet that is spanned by the composite Hilbert space $\mathcal{H}_{\text{ES}} = \mathcal{H}_{\text{orbit}} \otimes \mathcal{H}_{\text{spin}}$.
In the eigenstate basis of the orbital $\hat{\sigma}_z$ and spin $\hat{S}_z$ operators, the ES Hamiltonian reads:
\begin{align}\label{equ:Hes}
	%\begin{split}
		\hat{H}_{\text{ES}}/h &=
		D_{\text{ES}}^\parallel \hat{\mathbb{I}}_2 \otimes \left( \hat{S}_z^2 - \tfrac{2}{3} \, \hat{\mathbb{I}}_3 \right)
		-\lambda_{\text{ES}}^\parallel \hat{\sigma}_y \otimes \hat{S}_z \\ \nonumber
		+&D_{\text{ES}}^\perp \left[ \hat{\sigma}_z \otimes \left( \hat{S}_y^2 -\hat{S}_x^2 \right) - \hat{\sigma}_x \otimes \left( \hat{S}_y \hat{S}_x + \hat{S}_x \hat{S}_y \right)  \right] \\ \nonumber
		+&\lambda_{\text{ES}}^\perp \left[ \hat{\sigma}_z \otimes \left( \hat{S}_x \hat{S}_z + \hat{S}_z \hat{S}_x \right) - \hat{\sigma}_x \otimes \left( \hat{S}_y \hat{S}_z + \hat{S}_z \hat{S}_y \right)  \right] \\ \nonumber
		+&\muB g_{\text{ES}} \hat{\mathbb{I}}_2 \otimes \hat{\vec{S}} \cdot \vec{B} + \muB g_{l} B_z \hat{\sigma}_y \otimes \hat{\mathbb{I}}_3 \\ \nonumber
		+&d_{\text{ES}}^\perp \xi_x \hat{\sigma}_z \otimes \hat{\mathbb{I}}_3
		-d_{\text{ES}}^\perp \xi_y \hat{\sigma}_x \otimes \hat{\mathbb{I}}_3
		+d_{\text{ES}}^\parallel \xi_z \hat{\mathbb{I}}_2 \otimes \hat{\mathbb{I}}_3\,.
	%\end{split}
\end{align}
Here, $\hat{\sigma}_i$ are the Pauli matrices and $\hat{\mathbb{I}}_2$ the identity operator of the orbital subspace, and the vector $\vec{\xi}$ is the sum of the electric and strain-induced fields.  The $d_i$-factors are the respective components of the electric dipole moment. All other parameters are taken from the literature and are collected in Tab.~\ref{tab:1}.
Note that an axial strain $\xi_z$ only shifts the overall energy of the ES, which does not affect our analysis. For simplicity, we therefore will refer to $\dperp = d_{\text{ES}}^\perp \xi_\perp$ as the in-plane strain.

The above Hamiltonians are given in the eigenstate basis of the spin $\hat{S}_z$ and orbital $\hat{\sigma}_z$ operators.  We will refer to this basis as the \textit{$e_z$ basis}.
Its states are formed under dominating axial magnetic bias field $B_z$ and in-plane strain $\dperp$ along $\phi_\delta = 0$.  
The states are given by
\begin{align}\label{equ:Sz_state}
	%\begin{split}
		e_z &\textit{ basis} = \Bigl(\ket{+1}, \ket{0}, \ket{-1}, \\ \nonumber
		&\ket{E^0_{\x,+1}}, \ket{E^0_{\x,0}}, \ket{E^0_{\x,-1}}, \ket{E^0_{\y,+1}}, \ket{E^0_{\y,0}}, \ket{E^0_{\y,-1}}, \\ \nonumber
		&\ket{ss} \Bigr)\,,
	%\end{split}
\end{align}
where the first three states form the spin-triplet of the GS, the next six states form the orbital-doublet-spin-triplet of the ES, and the last state is the spin-singlet of the shelving state (SS). 
The SS consists of more than one level~\cite{doherty13} ($^1\text{A}_1$/$^1\text{E}$), but for our analysis, it is sufficient to model it by an effective singlet level.
$\x$ and $\y$ denote the two orbital branches $\Ex$ and $\Ey$ of the ES, which form under strain as shown in Fig.~\ref{fig:1}(a).

We will make use of two further sets of basis vectors.  For the situation of zero bias field and vanishing strain ($ \delta_\perp \to 0 $ at $ \phi_\delta = \SI{0}{\degree} $), the eigenvectors approximately assume the zero-field (zf) basis
\begin{align}\label{equ:zf_state}
	%\begin{split}
		\textit{zf basis} = \Bigl( &\ket{0}, \ket{-1}, \ket{+1}, \\ \nonumber
		&\ket{E_1}, \ket{E_2}, \ket{E^0_{\y,0}}, \ket{E^0_{\x,0}}, \ket{A_1}, \ket{A_2}, \\ \nonumber
		&\ket{ss} \Bigr)\,.
	%\end{split}
\end{align}
All ES states of the \textit{zf basis} have $\bigl \langle \hat{S}_z \bigr \rangle = 0$ and, apart from $\ket{E^0_{\x,0}}$ and $\ket{E^0_{\y,0}}$, also orbital eigenvalue $\expval{\hat{\sigma}_z}~=~0$.  While $\ket{E^0_{\y,0}}$ and $\ket{E^0_{\x,0}}$ are eigenstates to $\mS = 0$, the other eigenstates are linear combinations to equal parts of the $\mS = \pm 1$ and orbit $\x$ and $\y$ states.  The basis transformation $T_{\textit{zf}\rightarrow e_z}$ to the \textit{$e_z$ basis} is given in Appendix~\ref{app:basistrafo}.

Finally, under a large axial bias field $B_z$ and large strain $\dperp$ with angle $\phi_\delta \ne 0$, the eigenvectors are given by the high-field (hf) orbit and spin eigenbasis,
\begin{align}\label{equ:oe_state}
	%\begin{split}
		\textit{hf}&\textit{ basis} = \Bigl(\ket{+1}, \ket{0}, \ket{-1}, \\ \nonumber
		&\ket{E_{\x,+1}}, \ket{E_{\x,0}}, \ket{E_{\x,-1}}, \ket{E_{\y,+1}}, \ket{E_{\y,0}}, \ket{E_{\y,-1}}, \\ \nonumber
		&\ket{ss} \Bigr)\,.
	%\end{split}
\end{align}
The lower-lying branch is denoted by $\Ey$ and the upper-lying branch by $\Ex$. However, the axes no longer coincide with the $ x $ and $ y $ directions of the NV center coordinate system (Fig.~\ref{fig:0}), but rather with a coordinate system rotated by $ \phi_\delta $.  The \textit{hf basis} is shown in Fig.~\ref{fig:1}(c) and the basis transformation $T_{\textit{hf}\rightarrow e_z}$ to the \textit{$e_z$ basis} is given in Appendix~\ref{app:basistrafo}.

%%%%%
\subsection{Transition rates under optical excitation}
\label{sec:rates}

We next consider the transition rates between energy levels under optical excitation.
The associated rates are collected in Fig.~\ref{fig:1}(d) and Tab.~\ref{tab:1}.

Optical excitation from the GS to the ES can be achieved both resonantly (at cryogenic temperatures) and non-resonantly (at all temperatures).  If the excitation is non-resonant, it is followed by a rapid phonon relaxation.  
Both excitation methods are largely spin-preserving. Optical transitions with spin-flip occur with $\le\SI{2}{\percent} $~\cite{robledo11njp} and are therefore neglected here.

Since we aim to describe the population dynamics over a wide temperature range, we will focus on the non-resonant case. The resonant case could be covered by introducing individual excitation rates for each ES level.  While non-resonant excitation is in general not branch-selective, some selectivity can be achieved by adjusting the linear polarization of the laser~\cite{fu09, ulbricht16}. The linear polarizations in the $xy$-plane of the NV center are associated with the $\Ex$ and $\Ey$ orbital branches that are defined by the direction of the strain angle $ \phi_\delta $~\cite{batalov09, tamarat08}. Therefore, we introduce the excitation rates into $\Ex$ and $\Ey$ in the \textit{hf basis}.
Since these rates depend on the power $P$ of the laser, it is convenient to use the dimensionless optical saturation parameter $\beta = \beta_\x + \beta_\y$ and ratio $r_\beta = \beta_\x / \beta_\y$ as depicted in Fig.~\ref{fig:1}(d). A value of $\beta > 1$ indicates the onset of saturation~\cite{dreau11}.  To convert the laser power $P$ into the optical saturation parameter $\beta$, we define
\begin{equation}\label{equ:betaConv}
	\beta = P A
\end{equation}
with the alignment $A$ combining the effect of optical excitation efficiency and setup-specific conversion from laser power.
We assume a temperature-independent optical excitation efficiency and thus $ A $ as suggested by recent findings~\cite{blakley2023_arxiv}.

The radiative optical decay from the ES to the GS with rate $k_r$ is independent of the spin and the orbital branch and also largely spin-conserving.
The value of $k_r$ depends on the microscopic structure the NV center is embedded in. Typical bulk values are around $ \SI{67}{\per\micro\second} $~\cite{gupta16},
while those for NV centers in nanostructures, such as nanocrystals~\cite{reineck19} or nanopillars~\cite{ernst2023}, can be smaller.

The ES can also decay non-radiatively via the SS.  Opposite to the radiative $k_r$ decay, the decay rates from the six ES levels into the SS vary: The two levels associated with the $ \mS=0 $ spin states ($\ket{E^0_{\x,0}}$ and $\ket{E^0_{\y,0}}$) have the same and comparatively slow intersystem crossing (ISC) rate $ k_{E_\text{xy}} $.
The average ISC rate of the other four levels associated with the $ \mS=\pm 1 $ spin states is faster.
They vary for the different \textit{zf basis} states, as depicted in Fig.~\ref{fig:1}(d).
Following \citet{goldman15prb}, we use $ k_{A_{1}} = k_{E_{12}}/0.52 $ and $ k_{A_{2}} = 0 $ with $ k_{E_{1}} = k_{E_{2}} = k_{E_{12}}$ as given in Tab.~\ref{tab:1}.
These ISC rates have to be introduced to the model in the \textit{zf basis}.
By a basis transformation, we obtain the corresponding decay rates of the \textit{hf basis} states at high axial magnetic field and high strain $\dperp$. Levels with $ \mS=0 $ still decay with the same rate $ k_{E_\text{xy}} $. Since the \textit{hf basis} states with \ms{\pm1} are linear combinations to equal parts of \textit{zf basis} states with \ms{\pm1}, we find their decay rate by applying the $ \text{mean}\left( k_{E_{1}}, k_{E_{2}}, k_{A_{1}}, k_{A_{2}} \right) $, which evaluates to $\approx k_{E_{12}}$ for the values introduced above.
Likewise, the two rates $ k_{E_\text{xy}} $ and $ k_{E_{12}}$ are the ISC rates commonly used in room temperature models. 
We note that there is an additional temperature dependence of these ISC rates~\cite{goldman15prb, goldman_erratum_2017}. We neglect this effect as it is very small up to room temperature.

Finally, the decay from the spin-singlet SS back to the spin-triplet GS occurs with two characteristic rates $k_{S0}$ and $k_{S1}$ into the GS spin levels $\mS=0$ and $\mS=\pm 1$.
It is common to introduce the SS branching ratio as	$ r_S = k_{S0}/k_{S1} $.
For example, our branching ratio of $ r_S = 2.26 $~\cite{ernst2023} results in a \SI{70}{\percent} decay probability for $ m_s=0 $, and a \SI{15}{\percent} probability for each of the $ m_s = \pm 1 $ states.
We assume that $r_S$ is independent of temperature~\cite{thiering18}.
By contrast, the shelving state lifetime (SSL), defined by
\begin{equation}\label{equ:SSL}
	\tau_S (T) = \left(k_{S1}+k_{S0}\right)^{-1} = \tau_{S,0} \left( 1 - e^{-\frac{\Delta E}{\kB T}} \right)\,,
\end{equation}
varies with temperature $T$~\cite{robledo11njp, manson06, thiering18}.
This dependence results from a combination of spontaneous emission ($\tau_{S,0}$) and stimulated emission of a phonon with energy $\Delta E$ (see Tab.~\ref{tab:1}). $\kB$ is the Boltzmann constant.

Since $ \tau_S > \SI{100}{\nano \second} $ for all temperatures, the SS lifetime dominates the dynamics of the \ch{NV-} center under optical excitation together with $ 1/k_{E_{\text{xy}}} \approx \SI{100}{\nano \second} $.
An ES spin state with $ \mS=0 $ will dominantly decay optically and not take the path via the SS, maintaining its $ \mS $ value. A spin state $ \mS=\pm 1 $ on the other hand is likely to decay via the SS, which is a comparably slow process due to $ \tau_S $.  This selectivity leads to the generation of spin contrast $C$ in the photoluminescence (PL) intensity, discussed in Sec.~\ref{sec:ME} in Eq.~\ref{equ:C}.  
This selectivity further leads to a preferential population of the $\mS=0$ spin state after prolonged optical excitation, as we will discuss in Sec.~\ref{sec:SNR}.

%%%%%
\subsection{Phonon-induced transition rates}
\label{sec:phonon_rates}
Next, we consider phonon-driven population dynamics within the ES. The relevant effect for this work is phonon-mediated hopping between the two orbital branches.
This hopping arises from a coupling to the phonon bath where one- and two-phonon-processes (indices $1$ and $2$) drive transitions between the orbital states of the \ch{NV-} center. The temperature dependence of the thermal occupation of phonon modes dominates the temperature dependence of the \ch{NV-} photo-dynamics.
We start by introducing the up- and down rates
\begin{align}
	\Ey \rightarrow \Ex :\quad &k_{\uparrow}(T,\dperp, \eta) = k_{\uparrow,1} + k_{\uparrow,2}\,, \label{equ:khoppUp} \\
	\Ex \rightarrow \Ey :\quad &k_{\downarrow}(T,\dperp, \eta) = k_{\downarrow,1} + k_{\downarrow,2}\,, \label{equ:khoppDown}
\end{align}
respectively, which depend on the temperature $ T $ and the strain-induced splitting $ \hbar\Delta_\perp \approx 2h\dperp $ of the orbital branches (c.f. Fig.~\ref{fig:1}(b)). $ \eta $ parameterizes the coupling strength between electronic orbital states and phonons.
Importantly, $k_{\uparrow}$ and $k_{\downarrow}$ only act on the orbital subspace, leaving the state in the spin subspace untouched.

The rates are related via the Boltzmann distribution as
\begin{equation}\label{equ:detailedBallance}
	\frac{k_{\uparrow,1/2}}{k_{\downarrow,1/2}} = e^{-\frac{\hbar\Delta_\perp}{\kB T}} \,,
\end{equation}
due to spontaneous emission.
Derived in Appendix~\ref{app:phon}, the downwards one-phonon rate is given by 
\begin{equation}\label{equ:k1hoppDown}
	\begin{aligned}
		k_{\downarrow,1}(T, \dperp)
		&\approx 32 \eta h^3 \dperp^3 \left[ n(2 \dperp h) + 1\right]\,.
	\end{aligned}
\end{equation}
Note that the $ +1 $ term, associated with the spontaneous emission of a phonon, gives a finite and temperature-independent rate that needs to be considered even at $ T=\SI{0}{\kelvin} $ if strain is high.
For a two-phonon Raman process, we find
\begin{equation}\label{equ:k2T5}
	\begin{aligned}
		k_{\downarrow,2}(T,\dperp) \gtrapprox k_{\uparrow,2}(T,\dperp) = \frac{64\hbar}{\pi} \eta^2 \kB^5 T^5 I(T, \dperp)\,.
	\end{aligned}\\
\end{equation}
Here, $I(T, \dperp)$ is the integral over the phonon mode energies (in units of $\kB T$) in the Debye approximation
\begin{equation}\label{equ:I}
		 I(T, \dperp) = \int_{x_\perp}^{\Omega/\kB T} \frac{e^{x}x(x-x_\perp)\left[x^2+(x-x_\perp)^2\right]}{2 \left(e^{x}-1\right) \left(e^{x-x_\perp}-1\right)} dx\,,
\end{equation}
where $ x_\perp \approx 2\hbar\delta_\perp/(\kB T) $.
Note that $I(T, \dperp \rightarrow 0)$ is non-zero, meaning that two-phonon Raman transitions take place even in the absence of strain.
The integral is only mildly strain-dependent. Its temperature dependence is negligible if the cut-off energy is larger than the relevant temperature scales ($\Omega \gg \kB T$), which is the case in our work.
A thorough discussion of the effect of the cut-off energy is given in Sec.~\ref{sec:probing}.

Eqs.~\ref{equ:detailedBallance} -- \ref{equ:k2T5} indicate that $k_{\downarrow,1}$ will dominate for high strain at low temperatures. At elevated temperatures, up- and down rates converge and  $k_{\uparrow/\downarrow,2}$ will dominate due to their $T^5$-dependence. The temperature dependence of the hopping rates is further discussed and plotted in Sec.~\ref{sec:temp}.

%%%%% 
\subsection{Master equation}
\label{sec:ME}
We now discuss how we numerically simulate the population dynamics of the above rate model.
Because the phonon-induced transitions $k_{\uparrow/\downarrow}$ only act on the orbital subspace (leaving the coherent spin state undisturbed), dynamics cannot be simulated using classical rate equations commonly used at room temperature~\cite{tetienne12,robledo11njp} and recently also at low temperatures~\cite{happacher22}.  Instead, we use a master equation approach that acts on quantum states yet includes the known classical rates via jump operators.  We note that for high strain, the spontaneous emission rate $k_{\downarrow,1}$ can have a significant influence even at zero temperature~\cite{ernst2023}; therefore, a classical rate model might not be sufficient even in the low temperature limit.

To combine quantum states of spin and orbit with classical transition rates we use the master equation in Lindblad form~\cite{manzano20}.
The Liouville equation describes the time evolution of the density matrix $ \rho $ as
\begin{align}\label{equ:ME}
	%\begin{aligned}
			\dv{t} \hat{\rho} &= - \frac{i}{\hbar} \comm{\hat{H}}{\hat{\rho}} + \sum_{k} \left( \hat{L}_k \hat{\rho} \hat{L}_k^\dagger - \frac{1}{2} \acomm{\hat{L}_k^\dagger \hat{L}_k}{\hat{\rho}} \right) \\ \nonumber
			&\equiv \mathcal{L}(\hat{\rho})\,,
	%\end{aligned}
\end{align}
where $ \comm{\cdot}{\cdot} $ and $ \acomm{\cdot}{\cdot} $ are the commutator and anticommutator, $\hat{H}$ is the Hamiltonian, which we write as $ H = \text{diag}\left(H_{\text{GS}}, H_{\text{ES}}, 0\right) $ in matrix-form, and $ \mathcal{L} $ is the Liouvillian superoperator.
The first term in Eq.~\ref{equ:ME} is the von Neumann equation. The second term is the dissipator, which is also a superoperator.
The jump operators $ \hat{L}_k $ are required to introduce transition rates. A decay with Markovian statistics at rate $ k_{i,j} $ of basis state $ \ket{i} $ into basis state $ \ket{j} $ is associated with a jump operator
\begin{equation}\label{equ:Lk}
	\hat{L}_k = \sqrt{k_{i,j}} \ket{j}\bra{i}\,.
\end{equation}
Each decay channel in Fig.~\ref{fig:1}(d) is assigned an individual jump operator.
Examples in matrix-form are provided in Appendix~\ref{app:jump}. Importantly, the jump operators for the orbital branch hopping at rates $ k_{\downarrow/\uparrow} $ are constructed such that they preserve the coherent spin state.

We solve Eq.~\ref{equ:ME} in Fock-Liouville space (FLS) with the matrix $ \tilde{\mathcal{L}} $ and vector $ \vec{\rho} $ as
\begin{equation}\label{equ:rho}
	\vec{\rho}(t) = \exp(\tilde{\mathcal{L}}t)\vec{\rho}_0\,,
\end{equation}
since in our case, $ \mathcal{L} $ is time-independent (see Appendix~\ref{sec:impl} for details).
Note that parameters like the magnetic field and strain alter the Hamiltonian $ \hat{H} $ in Eq.~\ref{equ:ME} and thus lead to a different time evolution.
The steady state $ \rho_\infty  $ under given jump operators (e.g. under optical excitation) is given by the eigenvector of $ \tilde{\mathcal{L}} $ to eigenvalue $ 0 $.

%%%%% 
\subsection{Spin initialization and readout}
\label{sec:SNR}
The quantity observed in experiments is the $ \text{PL} $ intensity.
PL is generated by radiative emission (rate constant $k_r$) of the \ch{NV-} center.
In addition, contamination of the diamond surface by sp$^2$ carbon or adsorbed molecules, as well as temporary switching to the \ch{NV^0} charge state~\cite{bluvstein19} cause a fluorescent background.
We thus model the observed PL of the \ch{NV-} center as
\begin{equation}\label{equ:PL}
	\text{PL}(t) = A R \left(\sum_{i \in [4, 9]} \rho_{i,i} (t) k_r \right) + b P\,,
\end{equation}
where $ \rho_{i,i} $ are the state populations and  $ A $, $ R $, $ b $, and $ P $  are the setup-specific parameters as explained in Tab.~\ref{tab:1}.

To simulate $\text{PL}(t)$, we first construct the magnetic field- and strain-dependent Hamiltonian $H$ (Eqs.~\ref{equ:HgsNoStrain},\ref{equ:Hes}) and determine the \textit{hf basis} (Eq.~\ref{equ:oe_state}). Next, for every rate $k_i$, we construct the jump operators $L_{k_i}$ in the respective basis needed for that rate (Eq.~\ref{equ:Lk}). The temperature is required here to calculate the SSL (Eq.~\ref{equ:SSL}) and orbital hopping rate (Eq.~\ref{equ:khoppUp}, \ref{equ:khoppDown}).
We then transform the $L_{k_i}$'s to the same \textit{$e_z$ basis} as $H$ and solve the master equation for either the steady state $\rho_\infty$ or the time evolution $\rho(t)$ (Eq.~\ref{equ:rho}).
Finally, we calculate the $\text{PL}$ for the given setup parameters (Eq.~\ref{equ:PL}).

\begin{figure}
    \includegraphics{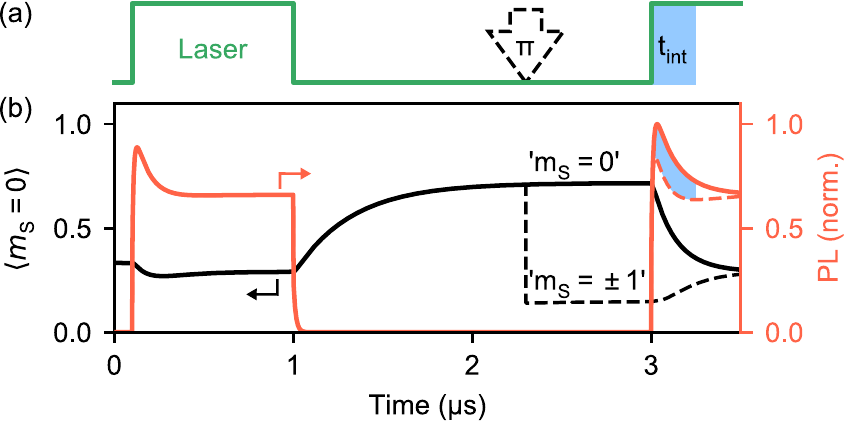}
	\caption{(a) Sequence of the initialization and readout protocol used to simulate the time $t$ resolved $\text{PL}(t)$ for determining the SNR. (b) Simulation of the expectation value $\left<\mS=0\right>$ (sum of GS and ES, black) and $\text{PL}(t)$ (orange).
    We start in a thermal state and obtain the spin state defined as the \ms{0} initialized state (solid lines) by a laser pulse, followed by a waiting time.
    The state defined as the \ms{\pm 1} initialized state (dashed lines) is obtained by an additional $ \pi $-pulse. In a second laser pulse, the integration time $ t_\text{int} $ for the spin-state readout is marked by the blue shaded area. The duration of the laser pulses and waiting time given here are typical experimental values \cite{ernst2023}. In simulations, we evaluate the steady-state solution for each step.
	} 
	\label{fig:pulseschemeSNR}
\end{figure}

An important figure-of-merit for quantum applications is the spin state readout fidelity, which can be quantified by the signal-to-noise ratio (SNR) of a single spin initialization and readout sequence~\cite{hopper18}.
To determine the SNR, we simulate the sequence depicted in Fig.~\ref{fig:pulseschemeSNR}(a):
First, we initialize the spin state by an off-resonant laser pulse, followed by a waiting time much longer than $ \tau_S $, until all populations have decayed to the GS.
The resulting spin state is defined as the \ms{0} initialized state. Note that due to $ k_{E_{xy}}, k_{S0} > 0 $, this state is a classical mixture of dominantly spin state $ \ket{0} $ and minor $ \ket{\pm 1} $ population and $\left<\mS=0\right> < 1$.
To obtain the \ms{-1}~($+1$) initialized state, we apply a state swap between the $ \ket{0} $ and $ \ket{-1} $ ($ \ket{+1} $) states.
Finally, we apply a second laser pulse and integrate the PL of the \ms{0} (\ms{\pm 1}) initialized state over a certain integration time $ t_\text{int} $ to obtain the average \PLms{0} (\PLms{\pm1}).
We then define the spin contrast $C$ (normalized blue shaded area in Fig.~\ref{fig:pulseschemeSNR}(b)) by
\begin{equation}\label{equ:C}
	C = 1 - \frac{\text{\PLms{\pm1}}}{\text{\PLms{0}}}
\end{equation}
and the SNR of a single readout~\cite{hopper18} by
\begin{equation}\label{equ:SNR}
	\text{SNR} = \sqrt{\text{\PLms{0}} \, t_{\text{int}}} \frac{C}{\sqrt{2-C}}\,.
\end{equation}
Note that this SNR accounts for both the spin-state initialization as well as readout fidelities~\cite{ernst2023}, which can be much reduced if spin-state relaxation is present in the ES, as discussed in Sec.~\ref{sec:implications_SNR}.
Further, the \ch{NV-} center's photo-dynamics change with temperature, magnetic field, and strain,  and therefore the $ t_\text{int} $ required for optimal SNR also changes.
In this work, we optimized $ t_\text{int} $ for best SNR in Eq.~\ref{equ:SNR} at the respective conditions. Since a faster spin relaxation in the ES reduces the optimal $ t_\text{int} $, a short $ t_\text{int} $ correlates with a low SNR (see Fig.~\ref{fig:init_read_tint}(b)).
We note that also the laser power could be optimized~\cite{wirtitsch23} for each condition, as discussed further in Sec.~\ref{sec:implications_SNR}. However, this optimization is beyond the scope of this work.

%%%%%%%%%%%%%%%%%%%%%%%
\section{Simulation of spin relaxation}
\label{sec:sim}
\subsection{Temperature dependence}
\label{sec:temp}

\begin{figure}
    \includegraphics{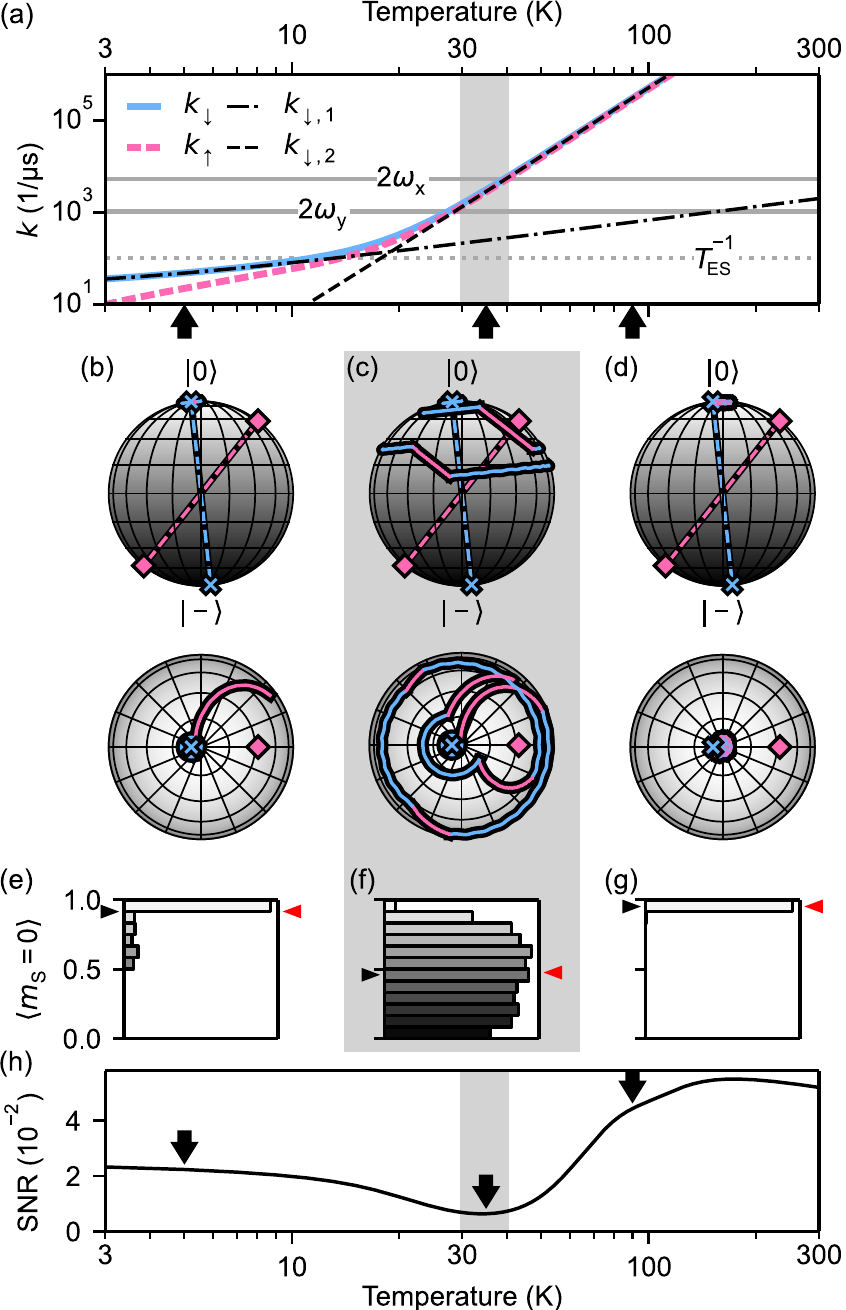}
	\caption{Illustration of the temperature-dependent spin relaxation process in the ES.
        We use the parameters from Tab.~\ref{tab:1}. In this setting, the ES levels are ordered as in Fig.~\ref{fig:1}(b).
        (a)~Orbital hopping rates and their one- and two-phonon contributions as a function of temperature (adapted from Ref.~\cite{ernst2023}). The inverse lifetime $ T_{\text{ES}}^{-1} \approx 1/\SI{10}{\nano\second} $ of the ES
        as well as the doubled Larmor frequencies $2\omega_{\x/\y}$ (units $\si{\mega\hertz}$) are indicated by horizontal lines.
		Spin relaxation is most efficient at temperatures where the hopping rates are resonant with $2\omega_{\x/\y}$, as indicated by the gray shaded area (here $\SIrange[range-phrase=-]{30}{40}{\kelvin}$).
		(b -- d) Exemplary realizations of $\SI{3}{\nano\second}$ time evolution on the side and top views of the Bloch sphere in the $\ket{0}$-$\ket{-}$-spin manifold, plotted for three selected temperatures (marked in (a) and (h)).
        The color of the time trace encodes the current orbital branch ($\Ex$: blue, $\Ey$: pink). Dashed lines show the quantization axes given by the energy eigenstates in spin space of the two orbital branches, about which Larmor precession occurs.
        (e -- g) Values of $m_{\mathrm{S}}$ after a time evolution of duration $ T_{\text{ES}} $ under random discrete jumps (average: black arrow) with a corresponding prediction of the master equation with Lindblad jump operators (red arrow).
		(h) Signal-to-noise ratio as a function of temperature. The three cases from above mark three distinct regimes, as discussed in the text. Prominently, the SNR reflects the rate of the spin relaxation process.
	} 
	\label{fig:2}
\end{figure}
The phonon-induced hopping between orbital branches (Sec.~\ref{sec:phonon_rates}) results in a strong temperature dependence of the ES spin dynamics. The interplay of orbital branch hopping (Eq.~\ref{equ:khoppUp}, \ref{equ:khoppDown}) and (temperature-independent) spin-state mixing in the ES Hamiltonian (Eq.~\ref{equ:Hes}) leads to a spin relaxation between the $\mS=0$ and $\mS=\pm 1$ states.
As an instructive example, we explain the spin relaxation process for the level structure given in Fig.~\ref{fig:1}(b) while assuming a strain of $ \delta_\perp = \SI{40}{\giga\hertz} $ (Tab.~\ref{tab:1}). 
In that setting, the state $ \ket{+} $ does not mix into the other spin eigenstates. Therefore, we depict the time evolution in the $\ket{0}$-$\ket{-}$-spin manifold and employ the Bloch sphere picture in Fig.~\ref{fig:2}. In our description, we write the mixed spin eigenstates as $\ket{0} - \varepsilon\ket{-}$ etc., where $\abs{\varepsilon}^2 = \abs{\epsilon_{\ket{\y}\ket{0}, \ket{\y}\ket{-}}}^2 \approx 0.1$ is a small spin-mixing amplitude. Here, $\epsilon_{\ket{i},\ket{j}}$ between other states $\ket{i}$ and $\ket{j}$ are negligibly small~\cite{ernst2023}.
External magnetic fields, as well as the in-plane strain angle $ \phi_\delta $ (here $ \phi_\delta = \SI{0}{\degree} $) and magnitude, alter the spin mixing compared to this example.
We will address these effects in Sections~\ref{sec:B}  and ~\ref{sec:strain}, while noting that the general mechanism described here remains valid. We further note that any spin mixing requires a finite $ \lambda_{\text{ES}}^\perp $~\cite{happacher2023, manson06, tamarat08}.

To illustrate the spin relaxation mechanism, we start with a Monte Carlo simulation of the time evolution of the spin state in the ES levels, which we will later relate to our full rate model. In these simulations, we exclusively model the coherent evolution in the ES and realize orbital transitions by random, discrete jump events with an average frequency given by $k_{\downarrow}$ and $k_{\uparrow}$. We start in the $ \ket{E_{\x,0}} $ state and evolve the spin state under the von Neumann equation while switching the spin Hamiltonian to fit the respective orbital branch at every jump.

The resulting time traces are shown in Fig.~\ref{fig:2}. The simulation demonstrates the importance of both the spin mixing amplitude $\varepsilon$ and the hopping rates $k_{\uparrow/\downarrow}$ (Fig.~\ref{fig:2}(a)) in the spin relaxation process: halting the Larmor precession in one orbital branch and continuing it with the same coherent spin state in the other orbital branch quickly moves the spin state away from the Bloch sphere pole where \ms{0}. For hopping rates $k_{\uparrow/\downarrow} \ll T_{\text{ES}}^{-1}$ slower than the ES lifetime, hopping events are rare and the spin state is mostly preserved. 

At  $5\unitformat{K}$, only the  $ k_{\downarrow,1} $ process leads to a rare decay to the $\Ey$ branch (Fig.~\ref{fig:2}(a)), which happens once in the exemplary time trace shown in Fig.~\ref{fig:2}(b).
At elevated temperature, hopping events become more frequent due to the rapidly increasing two-phonon Raman rates $k_{\uparrow/\downarrow,2}$.
Once
\begin{equation}\label{equ:2wisk}
    k_{\downarrow/\uparrow} \approx 2 \omega_{\x/\y} 
\end{equation}
we observe a resonance condition, where spin relaxation is most effective. Intuitively, we find that if precession about each quantization axis completes on average a one-half rotation, the initial spin state is relaxing most efficiently. In our setting, this occurs at around $35\unitformat{K}$ (Fig.~\ref{fig:2}(c)). 
Our effect is related to the $T_1$ relaxation process in a two-level system subjected to off-axis random telegraph noise (RTN). There, the shortest $T_1$ is found when $k = \pi \omega_0$~\cite{slichter90narrowing}.
In our example, the two-level system is the $\ket{0}$-$\ket{-}$-spin space, the RTN is a change in the Hamiltonian when alternating between the orbital branches, and the tilted quantization axis (due to $\varepsilon > 0 $) gives rise to the spin relaxation.
We note that the magnitude of $\lambda_{\text{ES}}^\perp$, which is still under debate~\cite{thiering18, doherty13}, does not affect the resonance condition described here.

For hopping rates $k_{\downarrow/\uparrow} \gg 2 \omega_{\x/\y}$ much faster than the Larmor precession frequency, the initial spin state is barely influenced. This is plotted at $ \SI{90}{\kelvin} $ in Fig.~\ref{fig:2}(d). The fast orbital dynamics allow for a description by an effective room-temperature Hamiltonian, which is obtained by performing a partial trace over the orbital subspace~\cite{plakhotnik14} of the low temperature Hamiltonian (dashed lines in Fig.~\ref{fig:1}(a,c) and Eq.~\ref{equ:HspinSubSpace}). In this description, $\ket{0}$ and $\ket{-}$ become eigenstates.
The orbital averaging is one reason why the \ch{NV-} center has such outstanding spin properties at room temperature.
This time-averaging process, which is analogous to motional narrowing~\cite{rogers09} in liquid-state NMR spectroscopy, has previously been addressed to explain the narrowing of the ES optically detected magnetic resonance (ODMR) lines at~\cite{fuchs10} and above room temperature~\cite{plakhotnik14, plakhotnik15}. In these studies, RTN along the $z$-axis of the \ch{NV-} center (i.e. $\omega_\x \ne \omega_\y$) gives rise to the orbital averaging and ES ODMR line narrowing. The spin relaxation process described in our work requires the presence of off-axis components in the RTN, which is caused by different spin mixing in the two branches.

Finally, we compare the Monte Carlo approach to our rate model (Eq. ~\ref{equ:ME}), which is used elsewhere in this work and in Ref.~\cite{ernst2023}.
In contrast to the discrete jumps in the Monte Carlo approach, the jump operators of the orbital hopping process in our rate model generate a continuous decay of the orbital state. To ensure that our rate model stays in the ES, all decay channels out of the ES were set to 0 for this comparison.
In Fig.~\ref{fig:2}(e -- g), we plot histograms of the the final population of $\mS = 0$ in Monte Carlo simulations after a time $T_{\text{ES}}$ spent in the ES, obtained from $10000$ ($\SI{5}{\kelvin}$, $\SI{35}{\kelvin}$) resp. $500$ ($\SI{90}{\kelvin}$) realizations of random time traces.
As expected, the average of the histogram yields the same result as the expectation value obtained from our rate model, validating both approaches.

\subsection{Implications for initialization and readout}
\label{sec:implications_SNR}

\begin{figure}
    \includegraphics{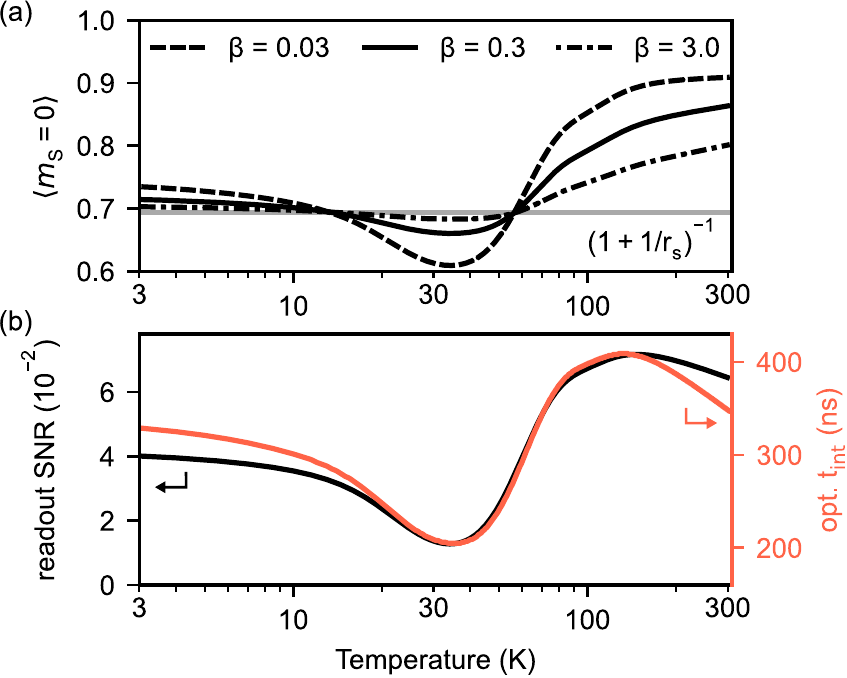}
	\caption{
        (a) Simulated spin initialization fidelity after a long laser pulse followed by a long waiting time (see Fig.~\ref{fig:pulseschemeSNR}) at different laser powers. $\beta = 0.3$ corresponds to the laser power $P = \SI{2.34}{mW}$ used elsewhere in this work. When spin relaxation is maximal (around \SI{35}{K}), increasing the laser power is beneficial, approaching a limit related to the SS branching ratio $r_S$ (horizontal grey line).
        (b) Simulated readout SNR (black), for which we assume an initialization of $\left<\mS=0\right> = 1$. We optimize the integration time $t_\text{int}$ (see Fig.~\ref{fig:pulseschemeSNR}) of the readout for each temperature. The resulting optimal $t_\text{int}$ (orange) follows the same trend as the readout SNR.
	} 
	\label{fig:init_read_tint}
\end{figure}

The temperature-dependent relaxation of the spin state in the ES has direct implications for quantum applications.
Common off-resonant spin initialization and readout schemes rely on a low spin-flip probability per optical cycle~\cite{robledo11njp, manson06}. However, as one can infer from Fig.~\ref{fig:pulseschemeSNR}(b), spin relaxation in the ES during laser excitation reduces $\left<\mS=0\right>$, independently affecting both the spin initialization and readout fidelities. In combination, this leads to a strong temperature dependence of the SNR, as plotted in Fig.~\ref{fig:2}(h).
Following the previous Sec.~\ref{sec:temp}, three regimes can be identified~\cite{ernst2023}:
(I) Around $ \SI{5}{\kelvin} $, the SNR is mostly constant and at a relatively high level.
(II) Around $ \SI{35}{\kelvin} $, the SNR is strongly reduced. The position of the minimum depends on the level spacing and thus Larmor frequencies of the given setting, while its depth additionally depends on the degree of spin mixing. We will discuss such dependencies in the next two sections.
(III) Around $ \SI{90}{\kelvin} $, the SNR has recovered and remains approximately constant up to room temperature. Here, best SNR is observed.

Our model allows us to study the individual contributions of the spin initialization and readout processes to the total suppression of SNR.
We focus on the resonance at $ k_{\downarrow/\uparrow} \approx 2 \omega $, where the spin relaxation erases most of the spin information in the ES. We first discuss the spin initialization fidelity in Fig.~\ref{fig:init_read_tint}(a).
In the limit of high laser power, the fidelity is determined by the decay of the SS as $\left<\mS=0\right> \approx (1+1/r_S)^{-1}$.
This limit arises from a predominant occupation of the SS before the initialization pulse ends (see Fig.~\ref{fig:pulseschemeSNR} and Ref.~\cite{manson06}).
When the laser power is decreased, the SS population and thus the initialization fidelity are reduced.
The lower limit of the fidelity is given by competing decay rates from the ES to the SS and the GS. The latter promotes a thermal spin population due to the ES spin relaxation.
% The latter causes a thermal spin population due to the loss of spin polarization in the ES.
Consequently, a higher laser power benefits the initialization fidelity around the resonance, which is in stark contrast to the situation at room temperature, where a lower laser power improves the initialization~\cite{wirtitsch23} (see Fig.~\ref{fig:init_read_tint}(a)). 
The reduction in the readout fidelity, on the other hand, only depends on the degree of the spin relaxation and typically dominates the temperature dependence of the overall SNR.
For the parameters used in this work, Fig.~\ref{fig:init_read_tint}(b) shows a drop of $\sim\SI{70}{\percent}$ in the readout performance at the resonance compared to the low-temperature limit. Conversely, the initialization fidelity is only reduced by $\sim\SI{15}{\percent}$ in Fig.~\ref{fig:init_read_tint}(a).
We note that in the above discussion, we neglect \ch{NV} center charge state dynamics, which can become significant at high laser power~\cite{wirtitsch23, aslam13}.

So far, we have discussed off-resonant spin initialization and readout schemes. If resonant excitation, which requires sufficiently narrow ($T\lesssim \SI{40}{\kelvin}$~\cite{fu09}) and stable spectral lines~\cite{chu14}, is available, one can use optical pumping to circumvent any spin-flip mechanism in the optical initialization cycle. This yields a higher $\left<\mS=0\right>$ initialization fidelity of $>\SI{99}{\percent}$~\cite{robledo11nature}. Here, a spin mixing in the ES is even beneficial for the duration of the initialization. However, single-shot resonant readout will still suffer from the spin relaxation in the ES ~\cite{robledo11nature, irber21}. 
Other readout schemes based on spin-to-charge conversion~\cite{shields15, hopper16, bourgeois15}, which are available at all temperatures, will equally suffer from the spin relaxation process.

We briefly address the implications on the charge state preparation of the NV center.
The initialization of the \ch{NV-} charge state is commonly achieved by green laser illumination~\cite{aslam13} (as in Fig.~\ref{fig:pulseschemeSNR}(a)). The temperature-dependent relaxation of the spin in the ES of the \ch{NV-} is expected to have a small influence on the initialization of the \ch{NV-} charge state:
Previous studies have found spin-dependent ionization rates from the ES~\cite{bourgeois15, shields15} and from the SS~\cite{happacher22, audecraik20, wolf23} of the \ch{NV-} to the neutral charge state \ch{NV^0} under green laser illumination. Since spin relaxation reduces the \ms{0} population, a temperature-dependent ionization rate to \ch{NV^0} is expected and was recently observed experimentally by \citet{blakley2023_arxiv}.
We neglect this small effect in our model.

%%%%%%%%%%%%%%%%%%%%%%%
\subsection{Magnetic field dependence}
\label{sec:B}
The temperature-dependent spin-state relaxation is influenced by external magnetic bias fields as well as the strength and direction of crystallographic strain or electric fields.
We first address the influence of an axial magnetic bias field. Two changes to the spin relaxation process can be identified:

(i) First, increasing the axial magnetic field leads to a dominating Zeeman term in the Hamiltonian and causes the spin states to be better eigenstates.
This reduces the spin mixing amplitudes $\epsilon_{\ket{i},\ket{j}}$ and, consequently, the strength of the spin relaxation. In the Bloch sphere picture of Fig.~\ref{fig:2}(c), this  corresponds to a smaller inclination angle of the eigenstate axis.

(ii) Second, since the Zeeman effect is linear in $B$, the level spacing and thus Larmor frequencies $\omega$ increase. Therefore, the condition $k_{\downarrow/\uparrow} \approx 2\omega$ for maximal spin relaxation shifts to a higher temperature. When only considering the dominant $ k_{\downarrow/\uparrow} \propto T^5$ dependence in Eq.~\ref{equ:k2T5}, the shift goes as $B^{1/5}$~\cite{happacher2023, ernst2023}.
Further, if the condition for maximal spin relaxation is met at higher temperature and smaller time scales, the relaxation rate at resonance increases. This is because the lifetime $ T_{\text{ES}} $ of the ES is constant while the number of hopping events increases.

As a result, the second effect (ii) partially compensates for the first (i). Therefore, even a very low spin mixing, as found at fields well above $\SI{1}{\tesla}$, can still result in an observable minimum in the SNR~\cite{ernst2023, happacher2023}.

Overall, applying a high axial magnetic field ($\gtrsim 100 \unitformat{mT}$) drastically improves SNR at low temperatures. At temperatures above $\gtrsim 50\unitformat{K}$, the SNR at high and low (few mT) magnetic fields becomes comparable \cite{ernst2023}.

%%%%%
\subsection{Strain dependence}
\label{sec:strain}
\begin{figure*}
	\includegraphics{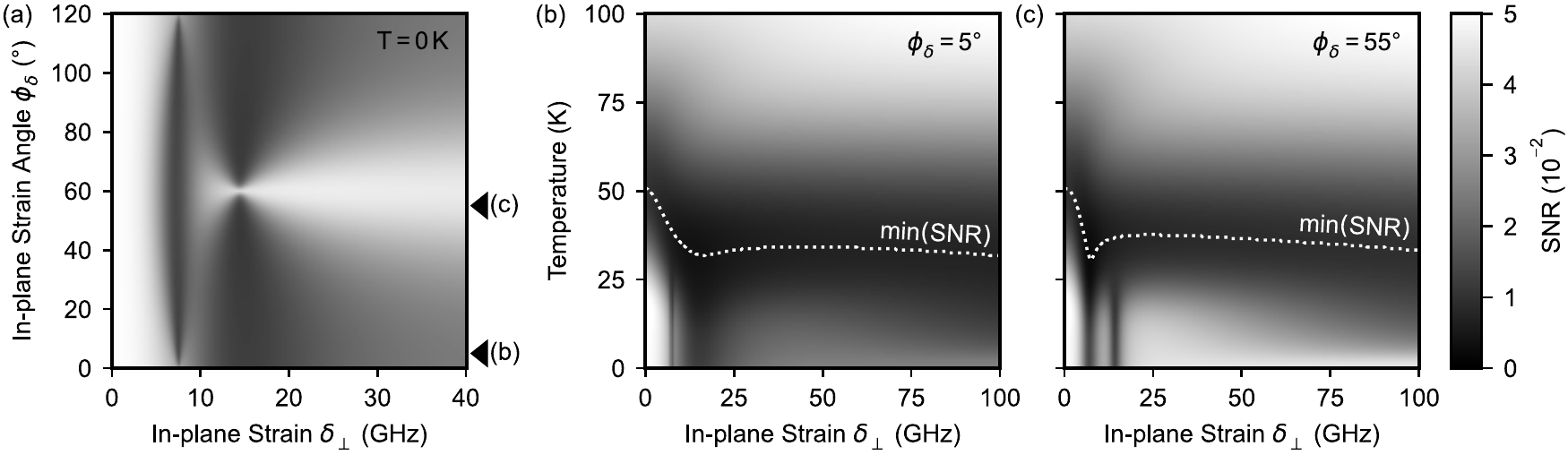}
	\caption{
		(a) Effect of strain $\delta_\perp$ or electric field and its angle $\phi_\delta$ on the SNR. ES LACs as indicated in Fig.~\ref{fig:1}(a) lead to regions of reduced SNR. (b, c) Effect of strain or electric field on the SNR as a function of temperature. The lines at $\SI{0}{\kelvin}$ for the two angles $ \phi_\delta $ in (b) and (c) correspond to the lines indicated by markers in (a). A multifaceted landscape of regions with reduced SNR is predicted by these simulations.
	} 
	\label{fig:3}
\end{figure*}
	
We next focus on the influence of strain or, equivalently, an electric field on ES spin relaxation.
We first discuss the effect of the in-plane angle $\phi_\delta$ for strain values $\delta_\perp \gtrsim 20\unitformat{GHz}$ at low temperature. The angle alters the degree of spin mixing in the two branches. 
In the $\Ey$ branch, the degree of spin mixing is maximal at $ \phi_\delta = \SI{0}{\degree} $ and vanishes toward  $ \phi_\delta = \SI{60}{\degree} $. In the $\Ex$ branch, this is exactly reversed and the spin mixing amplitudes are generally a fraction of the mixing that occurs in the $\Ey$ branch. If we assume equal optical excitation into both branches (i.e. $ r_\beta = 1$), this results in suppressed SNR at a strain angle of $ \phi_\delta = \SI{0}{\degree} $ (and multiples of $\SI{120}{\degree}$, due to the $C_{3 v}$ symmetry of the \ch{NV-} center), as plotted in  Fig.~\ref{fig:3}(a) for $T = 0\unitformat{K}$. The effect of $\phi_\delta$ disappears in the motional narrowing regime at elevated temperatures, as can be seen in Fig.~\ref{fig:3}(b-c).
We note that since this effect occurs at cryogenic temperatures, mitigation is possible by selective resonant excitation into the ES branch with the lower spin mixing at a given $\phi_\delta$~\cite{tamarat08}.

Next, we consider the effect of the strain magnitude $\delta_\perp$. Toward zero strain (flat eigenenergies in (Fig.~\ref{fig:1}(a))), as found in NV centers in bulk diamond, the spin state $\ket{0}$ has very small mixing with the spin states $\ket{\pm 1}$ ($ \abs{\epsilon_{\ket{i},\ket{j}}}^2 \le 0.03 $). This leads to optimal SNR at cryogenic temperature.
Since the level spacing and thus Larmor frequencies between eigenstates are large (see Fig.~\ref{fig:1}(a)), the maximal spin relaxation around $ k_{\downarrow/\uparrow} \approx 2 \omega $ is met at a temperature of around $ \SI{50}{\kelvin} $ (Fig.~\ref{fig:3}(b-c)). This constitutes an upward shift from the previously found minimum at $\SI{35}{\kelvin}$ at higher strain and smaller Larmor frequencies. Even at zero strain, however, there is still a well-pronounced minimum, caused by the same mechanisms (i) and (ii) explained in Sec.~\ref{sec:B} for the Zeeman effect at high magnetic fields.

Interestingly, this also implies that at temperatures above $\SI{50}{\kelvin}$, higher strain can be beneficial for the SNR since the averaging process is already effective. Below $\lesssim \SI{50}{\kelvin}$, low strained bulk NV centers will yield the best SNR.
In both figures Fig.~\ref{fig:3}(b-c), the effect of the strain-dependent one-phonon process (Eq.~\ref{equ:k1hoppDown}) at low temperatures~\cite{ernst2023} becomes apparent as a reduction of the SNR with increasing strain (compared at a constant $T \lesssim \SI{20}{K}$).
Overall, targeted application of electric fields~\cite{bassett11,tamarat08} or tuning of the strain magnitude and direction~\cite{lee16} can substantially improve the SNR.

%%%%%%%%%%%%%%%%%%%%%%%

\section{Comparison to experimental data}
\label{sec:experimental}

%%%%%
\subsection{Temperature dependence of PL intensity and spin contrast}
\label{sec:experimental_pl}
\begin{figure}
	\includegraphics{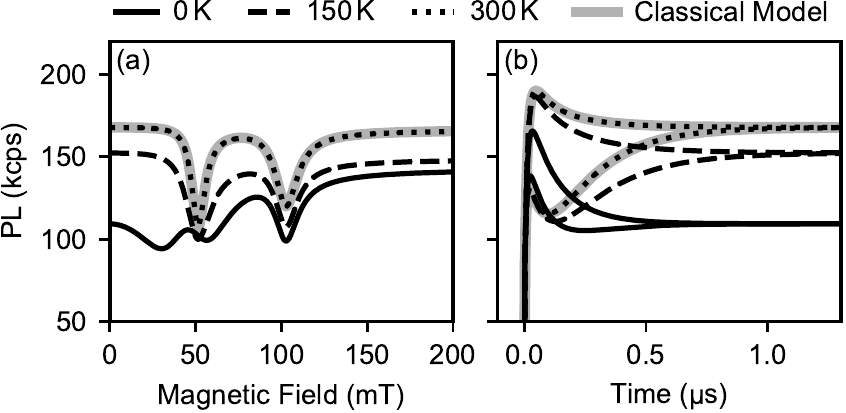}
	\caption{Comparison of our master equation model, evaluated at different temperatures, to the common rate model at $\SI{300}{\kelvin}$ (``Classical Model'').
    (a) Simulated steady state PL as a function of magnetic field. The two LACs of the ES average to a single LAC at room temperature (see Fig.~\ref{fig:1}(c) and Ref.~\cite{ernst2023}).  Here, we use $ \theta_B = \SI{1.9}{\degree} $ and $ \phi_B = \SI{194}{\degree} $ (NV-2 in Ref.~\cite{ernst2023}), and other parameters from Tab.~\ref{tab:1}.
    (b) $\text{PL}(t)$ dynamics during a readout laser pulse as in Fig.~\ref{fig:pulseschemeSNR}(b).
    In both panels, complete quantitative agreement with the classical rate model is recovered at room temperature.
	} 
	\label{fig:comparisonToClassical}
\end{figure}
Based on the same parameters in Tab.~\ref{tab:1} and the discussion in Sec.~\ref{sec:rates} we compare our master equation model with the established rate model at room temperature~\cite{robledo11njp, tetienne12}.
In Fig.~\ref{fig:comparisonToClassical}, we verify that our model, which is based on the zero temperature Hamiltonian and rates, recovers complete quantitative agreement with the classical rate model at room temperature.
Further, we simulate the temperature dependence of PL and spin contrast and compare it with an extensive set of experimental data. We find excellent agreement between simulation and experiment over a wide parameter range in Ref.~\cite{ernst2023}.
Here, we will further verify the predictions of our model by comparing it to previous work.

%%%%%
\subsection{Intersystem crossing rates at cryogenic temperatures}
\label{sec:comparisonISC}
\begin{figure}%[t]
	\centering
	\includegraphics{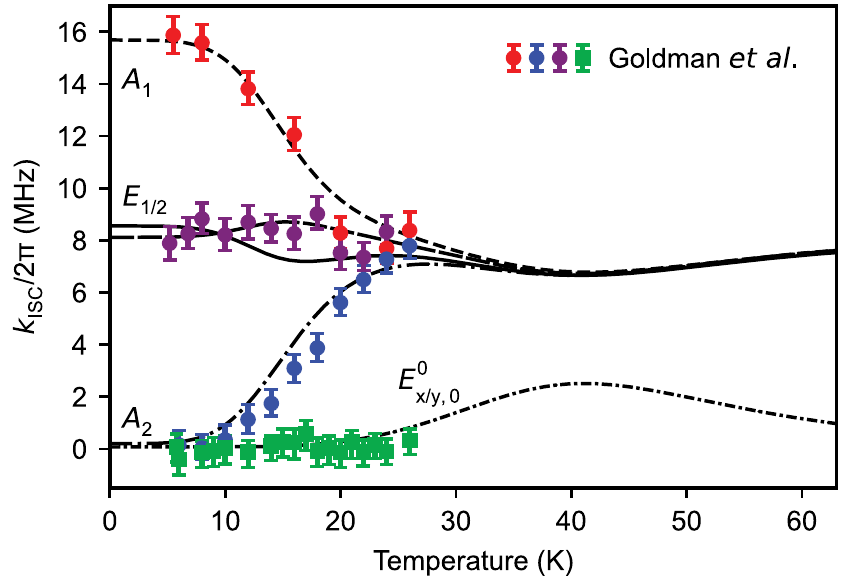}
	\caption{Simulation of effective ISC rates obtained from a single exponential decay fit of the PL after exciting into the various states of the ES in conditions where \textit{zf basis} states are almost eigenstates (labels). Experimental data and parameters are taken from \citet{goldman15}.
	}
	\label{fig:ISC}
\end{figure}

In Fig.~\ref{fig:ISC} we plot the effective ISC rates $k_{\text{ISC}}$ for the different ES levels of the \textit{zf basis} as measured by \citet{goldman15}. In their work, resonant laser excitation into the individual low strain eigenstates was used in the temperature range $\SIrange[range-phrase=-]{5}{26}{\kelvin}$.
Probing the ISC rates at higher temperatures is experimentally not possible by resonant laser excitation due to the broadening of the spectral lines. We find remarkable agreement between our model and data from Ref.~\cite{goldman15} (for implementation details, see Appendix~\ref{sec:SIcomparisonISC}). First, we observe that the ISC rates with the same $ \abs{m_{\mathrm{S}}} $ average with increasing temperature, as reported previously~\cite{goldman15}.
Second, in the region of maximal spin relaxation, our simulation shows an increase of the effective ISC rate of the $ m_{\mathrm{S}}=0 $ states and the corresponding decrease of the $ \abs{m_{\mathrm{S}}}=1 $ states.
We note that the different ISC rates below $ \SI{20}{\kelvin} $ can only result from a model where rates are introduced in \textit{zf basis}, as discussed in Sec.~\ref{sec:rates}
(independent of whether a classical rate model or a master equation model is used).

%%%%%
\subsection{Strain dependence at room temperature}
\label{sec:trf}
\begin{figure}%[t]
	\centering
    \includegraphics{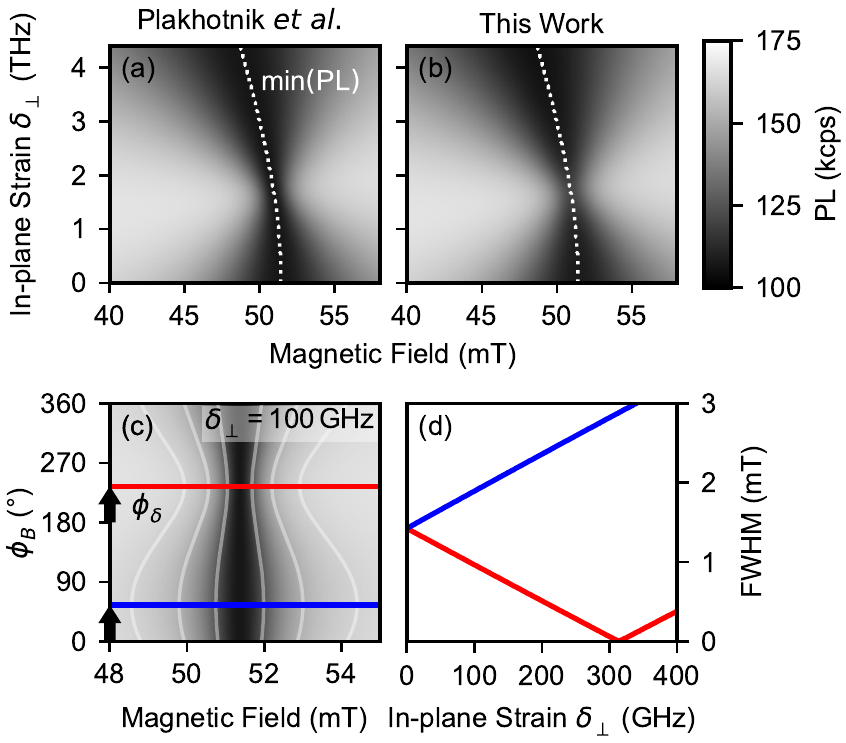}
	\caption{Simulated steady state $ \text{PL} $ at the ES LAC as a function of magnetic field and strain at $\SI{300}{\kelvin}$.
		(a) Room temperature classical rate model with the additional temperature reduction factor in the Hamiltonian as derived by \citet{plakhotnik14}. The LAC (minimum in $ \text{PL} $) is marked by the dotted line. A magnetic field alignment of $ \theta_B = \SI{1.9}{\degree} $ and $ \phi_B = \SI{194}{\degree} $ (NV-2 in Ref.~\cite{ernst2023}, other parameters as in Tab.~\ref{tab:1}) is used. 
		(b) The master equation model developed in this work. The same line plotted in (b) is shown to demonstrate the excellent overlap with the predicted position of the LAC.
        (c) For a given magnetic field alignment angle $\theta_B$ (here: $\SI{0.3}{\degree} $), a sweep of the in-plane magnetic field angle $ \phi_B $ is predicted to reveal the strain angle $ \phi_\delta $ and magnitude $\delta_\perp$. White lines are contours of constant $ \text{PL} $.
        (d) Full width at half maximum (FWHM) of the LAC at the narrowest ($\phi_B = \phi_\delta + \SI{180}{\degree}$, red line) and widest ($\phi_B = \phi_\delta$, blue line) in-plane magnetic field angle $ \phi_B $ as a function of $\delta_\perp $. Magnetic field sweeps to determine the FWHM are indicated in (c) for one value of $\delta_\perp $.
	}
	\label{fig:trfmap}
\end{figure}

In Sec.~\ref{sec:strain} we discussed that the strain dependence of the ES disappears towards room temperature.
However, previous studies found a remaining strain dependence of the ES at room temperature.
Initially, an additional empirical term was introduced to the room temperature model~\cite{fuchs08, neumann09} to account for observations in highly strained NV centers, as found in nanodiamonds.
It was then shown analytically by \citet{plakhotnik14} that this term arises from a sufficiently large imbalance in the Boltzmann-distributed (Eq.~\ref{equ:detailedBallance}) ES branch population, caused by the large branch-splitting at very high strain ($2\dperp \approx \SI{1}{\tera\hertz}$ in their work). This prevents the orbital averaging process from completing and thus leads to a remaining influence of the low temperature strain dependence even above room temperature.

To compare our model with the experimentally verified model by Plakhotnik \etal, we simulate the strain dependence of the averaged ES LAC around $\SI{51}{\milli\tesla}$ (see Fig.~\ref{fig:1}(c)).
As discussed in Refs.~\cite{happacher22, ernst2023}, LACs of the GS and time-averaged ES cause a strong reduction in the $ \text{PL} $. The LAC position gives insight into the level structure and can be used to compare the levels of a room temperature model with the numerically time-averaged levels as predicted by our model.
To that end, we extend the classical rate model by Plakhotnik \etal{} to take applied magnetic fields into account (details in Appendix~\ref{sec:SItrf}).

In Fig.~\ref{fig:trfmap}(a,b), we compare the two models.
We first recall that the accepted room temperature rate model does not contain an explicit dependence on crystal strain in the ES (see dashed lines in Fig.~\ref{fig:1}(a)) --
orbital averaging removes any strain dependence of the ES.
However, for the high strain values in Fig.~\ref{fig:trfmap}(a), a classical model that includes the Boltzmann distribution predicts a strain dependence of the ES LAC.
Likewise, in Fig.~\ref{fig:trfmap}(b) we plot the master equation model developed in this work.
We find quantitative agreement between the two models.
Our model, which includes the Boltzmann distribution using different up- and downward hopping rates (Eq.~\ref{equ:detailedBallance}), readily predicts the behavior observed under extreme strain at room temperature. This was previously seen as a central challenge for a model covering the entire temperature range~\cite{doherty13}.

Remarkably, both models predict a non-trivial strain dependence of the width of the LAC, as plotted in Fig.~\ref{fig:trfmap}(a,b). The width of the LAC as a function of magnetic field varies with the in-plane strain direction $\phi_\delta$, and the strain position of the smallest width shifts with the magnetic field direction $\theta_B$ and $\phi_B$.
This field dependence could be employed to characterize the order of magnitude of crystal strain and its direction.
Since strain strongly affects the SNR of the \ch{NV-} center at low temperatures (see Fig.~\ref{fig:3}), this feature could be used for all-optical pre-characterization at room temperature.
As simulated with the analytical model by Plakhotnik \etal{} in Fig.~\ref{fig:trfmap}(c,d), elevated strain leads to a $\delta_\perp$-characteristic narrowing of the LAC at an in-plane magnetic field angle $\phi_B = \SI{180}{\degree}+\phi_\delta$.

%%%%%%%%%%%%%%%%%%%%%%%
\subsection{Probing of spin-phonon interactions}
\label{sec:probing}

\begin{figure}
    \includegraphics{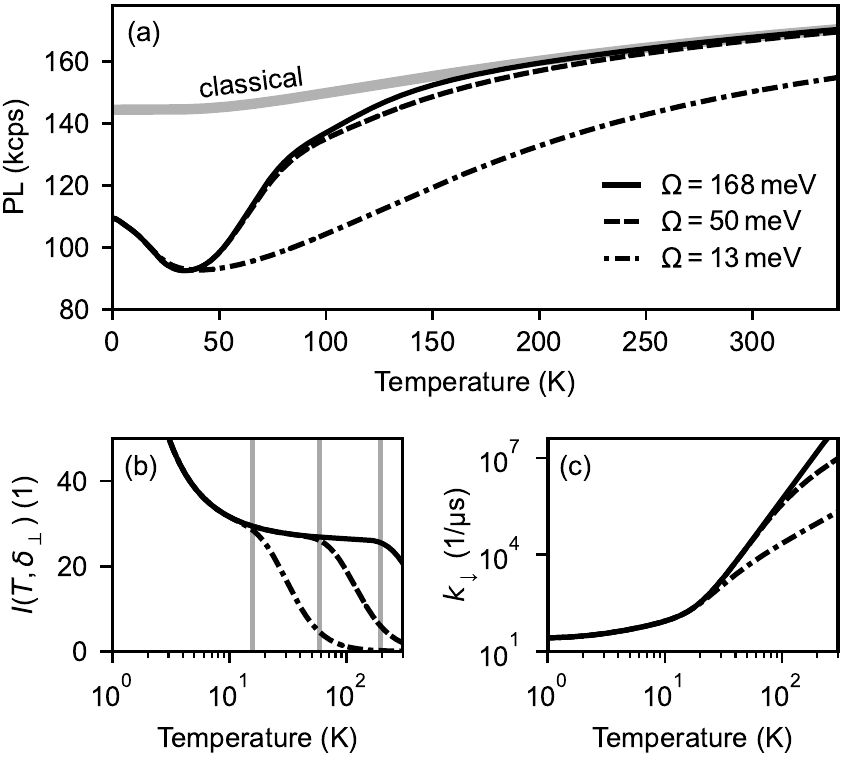}
	\caption{(a) Steady-state PL as a function of temperature for several phonon cutoff energies $\Omega$. The $ \Omega $ values correspond to the discussion in Sec.~\ref{sec:probing}. The increase in PL above $\SI{100}{\kelvin}$ is due to the decreasing SSL (Eq.~\ref{equ:SSL}). The effect of the SSL can be distinguished from the ES phonon-driven averaging by comparison with the classical room temperature model (gray curve).
    (b) Phonon mode integral  $I(T,\dperp)$ plotted for the different $\Omega$ values. Vertical gray lines indicate the onset of the cut-off at $T \approx 0.1\cdot\Omega/\kB$.
    (c) $k_{\downarrow}$ plotted for the different $\Omega$ values.
    } 
	\label{fig:4}
\end{figure}

The spin relaxation process discussed above is dependent on the number of available phonons in thermally activated modes and the strength of their coupling to the ES orbital states. Characterizing this spin relaxation process thus suggests a new tool for investigating the interaction of the NV center with the phonon bath.
As the temperature dependence of the orbital hopping rate is still under debate~\cite{gali19}, we will put recent experimental findings
into context with previous studies
and show how our model allows probing electron-phonon interactions based on simple PL measurements.

Multiple studies have determined the electron-coupling strength $\eta$ and reported similar values~\cite{ernst2023, abtew11, plakhotnik15, goldman15, goldman15prb, goldman_erratum_2017, happacher2023}. Determining $\eta$ requires the evaluation of the integral over the phonon spectrum $I(T,\dperp)$ for the two-phonon Raman process in Eq.~\ref{equ:k2T5}, which all of those studies solve in the Debye approximation. However, different studies use  different phonon cut-off energies. We consider the following studies:

(I) Our related work~\cite{ernst2023} reports measurements of the ODMR contrast and PL intensity in a temperature range from $T = \SI{3}{\K}$ -- $\SI{300}{\K}$. We assume the Debye energy of diamond $\Omega = 168 \unitformat{meV}$ for the cut-off.

(II) \citet{goldman15} uses resonant photoluminescence excitation (PLE)  spectroscopy to probe the ES dynamics up to $\SI{26}{\kelvin}$. At higher temperatures, the spectral broadening does not allow the use of PLE spectroscopy. They use $\Omega = \infty $.

(III) \citet{abtew11} determine the cut-off energy by fitting measurements~\cite{fu09} of the line width of the zero phonon line (ZPL) from $T = \SI{3}{\K}$ -- $\SI{250}{\K}$ and obtain $\Omega=50 \unitformat{meV}$.

(IV) \citet{plakhotnik15} perform measurements of the ES ODMR line width in the temperature range $\SIrange[range-phrase=-]{300}{550}{\kelvin}$. Since their measurements rely on the time-averaged ES, they extrapolate their model to cryogenic temperatures. They fit a cut-off energy of $\Omega = 13.4 \unitformat{meV}$. 

We test the different cut-off energies by inspecting their effect on our rate model. First, in Fig.~\ref{fig:4}(a), we simulate the temperature-dependent PL to show that there is a negligible difference between our cut-off ($168 \unitformat{meV}$) and the lower cut-off from Abtew \etal{} ($50 \unitformat{meV}$). This is because the time-averaging sets in at much lower energies ($\kB T \sim 3\unitformat{meV}$ for $T = 35 \unitformat{K}$) than both cut-offs. In extension, no discernible effect arises when using $\Omega \rightarrow \infty $.

By contrast, the low cut-off value of $\Omega = 13.4 \unitformat{meV}$ from Plakhotnik \etal{} results in a slower increase in the hopping rates toward high temperature (Fig.~\ref{fig:4}(c)). Consequently, the recovery of the PL does not complete up to room temperature (Fig.~\ref{fig:4}(a)). This is in contradiction with several recent experimental studies \cite{ernst2023, happacher2023, blakley2023_arxiv}, as well as initial work by \citet{rogers09}, which found in good agreement that the averaging process and thus the PL recovery completes around $\SI{100}{\kelvin}$. 
Plakhotnik \etal{} explain the much higher value of $\Omega$ found by Abtew \etal{} for the ZPL line width by contributions of $\text{A}_1$-symmetric phonon modes. However, this does not apply to the orbital averaging underlying our rate model, which is caused by E-symmetric phonons. To gain further insight, we expect that our model, coupled with a detailed phonon spectrum as well as a PL or spin contrast measurement in the range $\SIrange{80}{150}{\kelvin}$, has to be used.

%%%%%%%%%%%%%%%%%%%%%%%
\section{Conclusions}
\label{sec:conclusion}
In summary, we present a master-equation model of the \ch{NV-} center population dynamics that unifies the existing rate models developed for the low and high temperature limits.  We include the effect of temperature by introducing phonon-induced hopping between the ES energy levels, which, together with spin mixing, is the key mechanism for spin-state relaxation in the ES.  The relaxation process is most effective when the hopping rate between the orbital states is resonant with the ES spin level spacing, explaining the reduction in the PL intensity at intermediate temperatures.  Since the ES level spacing and mixing are dependent on strain and magnetic field, the resonance condition can be tuned by these parameters.  Using numerical simulations of the population dynamics, we extract several important experimental observables, including the dynamic and steady-state PL intensity, as well as the SNR for spin-state readout, which is relevant for quantum applications.

We further show that by systematic modeling of the PL intensity as a function of temperature, we can probe electron-phonon interactions.  This approach is applicable in regimes where resonant laser PL excitation spectroscopy~\cite{goldman15} and measurements of motional narrowing on ES ODMR lines~\cite{plakhotnik15} are unavailable, providing a new avenue for experimental probing of the contributing phonon modes.  Applying our analysis to recent experimental data~\cite{ernst2023, happacher2023, blakley2023_arxiv}, we find inconsistencies between the experimental observation and the present theoretical understanding~\cite{gali19} of the spin-phonon interaction.

%%%%%%%%%%%%%%%%%%%%%%%
\section*{Acknowledgments}

The authors thank Assaf Hamo, Konstantin Herb, William Huxter, Dominik Irber, Fedor Jelezko, Luca Lorenzelli, Patrick Maletinsky, Francesco Poggiali, Friedemann Reinhard, J\"org Wrachtrup and Jonathan Zopes for useful input and discussions.
This work was supported by the European Research Council through ERC CoG 817720 (IMAGINE), the Swiss National Science Foundation (SNSF) through Project Grant No. 200020\_175600 and through the NCCR QSIT, a National Centre of Competence in Research in Quantum Science and Technology, Grant No. 51NF40-185902, and the Advancing Science and TEchnology thRough dIamond Quantum Sensing (ASTERIQS) program, Grant No. 820394, of the European Commission.

%%%%%%%%%%%%%%%%%%%%%%%
\appendix

\section{Basis Transformations}
\label{app:basistrafo}
The transformation for a basis change from \textit{zf} to \textit{$e_z$ basis} of the matrix representation of an operator $ \hat{M} $ is~\cite{doherty11}
\begin{equation}\label{equ:trafo}
	M_{e_z} = T_{\textit{zf} \rightarrow e_z} M_{\textit{zf}} \left(T_{\textit{zf} \rightarrow e_z}\right)^{-1}\,,
\end{equation}
with
\begin{equation}\label{equ:TzfSz}
	T_{\textit{zf} \rightarrow e_z} = \mqty(
	0 &    0 &    1 &     0 &    0 &    0 &    0 &    0 &    0 &    0 \\
	1 &    0 &    0 &     0 &    0 &    0 &    0 &    0 &    0 &    0 \\
	0 &    1 &    0 &     0 &    0 &    0 &    0 &    0 &    0 &    0 \\
	0 &    0 &    0 &  -i/2 &  1/2 &    0 &    0 & -i/2 & -1/2 &    0 \\
	0 &    0 &    0 &     0 &    0 &    0 &    1 &    0 &    0 &    0 \\
	0 &    0 &    0 &  -i/2 & -1/2 &    0 &    0 & -i/2 &  1/2 &    0 \\
	0 &    0 &    0 &   1/2 &  i/2 &    0 &    0 & -1/2 &  i/2 &    0 \\
	0 &    0 &    0 &     0 &    0 &    1 &    0 &    0 &    0 &    0 \\
	0 &    0 &    0 &  -1/2 &  i/2 &    0 &    0 &  1/2 &  i/2 &    0 \\
	0 &    0 &    0 &     0 &    0 &    0 &    0 &    0 &    0 &    1 \\
	)\,.
\end{equation}

Likewise, the transformation matrix to the \textit{$e_z$ basis} from the \textit{hf basis} analog to Eq.~\ref{equ:trafo} can be constructed as
\begin{equation}\label{equ:ToeSz}
	T_{\textit{hf} \rightarrow e_z} = \text{diag}\left(\mathbb{I}_3, T^{\text{ES}}_{\textit{hf} \rightarrow e_z}, 1\right)\,,
\end{equation}
with $ T^{\text{ES}}_{\textit{hf} \rightarrow e_z} = T^{\text{orbit}}_{\textit{hf} \rightarrow e_z} \otimes \mathbb{I}_3 $. $ T^{\text{orbit}}_{\textit{hf} \rightarrow e_z} $ is construed column-wise from the eigenvectors written in \textit{$e_z$ basis} of the Hamiltonian in the orbital subspace with
\begin{equation}\label{equ:HorbSubSpace}
	H_{\text{orbit}} = 1/3 \cdot \Tr_{\text{spin}}\left( H_{\text{ES}} \right)
\end{equation}
which can be obtained from Eq.~\ref{equ:Hes} by the partial trace over the spin subspace.
Likewise, we define the Hamiltonian in the spin subspace with
\begin{equation}\label{equ:HspinSubSpace}
	H_{\text{spin}} = 1/2 \cdot \Tr_{\text{orbit}}\left( H_{\text{ES}} \right) \,.
\end{equation}

\section{Derivation of phonon-induced orbital hopping}
\label{app:phon}
We now derive the orbital hopping rate that is introduced to the orbital electronic state of the \ch{NV-} ES via coupling to the phonon bath.
In our derivation, we closely follow previous work by \citet{walker68} that was first adapted to the \ch{NV-} center by \citet{fu09} and further developed by Abtew \etal~\cite{abtew11}, Goldman \etal~\cite{goldman15prb}, and Plakhotnik \etal~\cite{plakhotnik15, plakhotnik14}.
For a recent review of the topic in the context of vibronic states of the \ch{NV-} center we refer the reader to Ref.~\cite{gali19}.
But here, in contrast, we aim to provide all steps of the derivation which allows us to understand how all terms arise. Also, in contrast to the high temperature and high strain approximations in~\cite{plakhotnik15} or the low temperature and low strain approximations in~\cite{goldman15prb}, we will keep the full equations allowing us to model the wide range of parameters covered in this work.
In the following, index $ i $ should denote the $ i^\text{th} $ vibrational mode. $ p_i \in \{\x,\y\} $ be the polarization of that mode.
States $ \ket{O} \in \{\ket{\x},\ket{\y}\} $ are orbital eigenstates (Eq.~\ref{equ:oe_state}).
A vibrational level is $ \ket{O}\ket{\chi_m} $ with $ \ket{\chi_m} = \prod_{i} \ket{n_i} $ and energy $ \epsilon_m = \sum_{i} n_i \omega_i \hbar $. $ m $ denotes the set of occupation numbers $ \{n_i\} $ of modes $ i $ (in contrast to the Larmor frequencies $\omega_{\x/\y}$ in the main text, these $\omega_i$ are in units $\si{\rad/\second}$).
Under the influence of the dynamic Jahn-Teller effect, the electron-phonon interaction with E-symmetric phonons of the ES is~\cite{fu09}
\begin{align}\label{equ:Hep}
	%\begin{split}
		H_{\text{e-p}} = \sum_{i} [ &\delta_{p_i,\x} \hbar \lambda_{i} (\op{\x}{\x} - \op{\y}{\y}) (a_i + a_i^\dagger) \\ \nonumber
		&-\delta_{p_i,\y} \hbar \lambda_{i} (\op{\x}{\y} + \op{\y}{\x}) (a_i + a_i^\dagger) ]\,.
	%\end{split}
\end{align}
The effect of the annihilation $ a_i $ and creation $ a_i^\dagger $ operators for mode $ i $ on phonon states is denoted as:
\begin{align}
	a_i \ket{\chi_m} &= \sqrt{n_{m,i}} \ket{\chi_{m,i-}} \label{equ:annihil}\\
	a_i^\dagger \ket{\chi_m} &= \sqrt{n_{m,i}+1} \ket{\chi_{m,i+}} \label{equ:creation}
\end{align}
For the derivation, we will assume to start in a state $ \ket{s} $ of the lower branch $ \Ey $ and find the rate with which it decays to a final state $ \ket{f} $ of the upper branch $ \Ex $. Thus,
\begin{align}
	\ket{s} &= \ket{\y}\ket{\chi_s} \label{equ:start}\\
	\ket{f} &= \ket{\x}\ket{\chi_f} \label{equ:end}
\end{align}
and the energy splitting (final minus initial) between the orbital electronic states is given by
\begin{equation}\label{equ:splitting}
	\hbar \Delta_\perp = \ev{H_{\text{orbit}}}{\x} - \ev{H_{\text{orbit}}}{\y} \approx 2 \dperp h > 0\,,
\end{equation}
with the eigenenergies of the orbital subspace $ H_{\text{orbit}} $ from Eq.~\ref{equ:HorbSubSpace}. The approximation holds as long as $ g_l B_z $ is small, which is usually the case but recent observations suggest a larger $  g_l $ in high strained NV centers at $ B_z \gg \SI{1}{\tesla} $~\cite{happacher22}.
The transition rate between the two orbital branches can be calculated by Fermi's Golden Rule
\begin{equation}\label{equ:Fermi}
	k_{\uparrow} = \frac{2\pi}{\hbar} \sum_{f}  \abs{T_{fs}}^2 \delta(\hbar\Delta_\perp+\epsilon_f-\epsilon_s)\,,
\end{equation}
where $ T_{fs} $ is the transition matrix element.
Up to second order, $ T_{fs} $ can be found from
\begin{align}
	T_{fs} &= T_{fs}^{(1)} + T_{fs}^{(2)} + ... \label{equ:Tfs}\\
	T_{fs}^{(1)} &= \mel{f}{H_{\text{e-p}}}{s} \label{equ:Tfs1}\\
	T_{fs}^{(2)} & = \sum_{m} \begin{aligned}[t] &\frac{1}{E_s-E_m} \\
			&\cdot \bra{\chi_f}\mel{\x}{H_{\text{e-p}}}{O_m}\ket{\chi_m} \\
			&\cdot \bra{\chi_m}\mel{O_m}{H_{\text{e-p}}}{\y}\ket{\chi_s} \\
		\end{aligned}\label{equ:Tfs2}
\end{align}
with the sum over all possible intermediate states $ \ket{O_m}\ket{\chi_m} $ with energy $ E_m $ for the second order process.

\subsection{One phonon process}
\label{app:1phon}
First, we derive the first-order transition matrix element, which is a one-phonon process. The energy gap $ \hbar \Delta_\perp $ is overcome by the absorption of one phonon with exactly that energy.
\begin{align}\label{equ:Tfs1_a_adagg}
	T_{fs}^{(1)} &= \mel{f}{H_{\text{e-p}}}{s} \\ \nonumber
	&= -\sum_{i}\delta_{p_i,\y}\hbar\lambda_i \left(\mel{\chi_f}{a_i}{\chi_s}+\mel{\chi_f}{a_i^\dagger}{\chi_s}\right) \\ \nonumber
	&= -\sum_{i}\delta_{p_i,\y}\hbar\lambda_i \left(\sqrt{n_{s,i}}\delta_{f,si-}+\sqrt{n_{s,i}+1}\delta_{f,si+}\right)
\end{align}
Here, $ \delta_{f,si-} $ denotes that $ \ket{\chi_s} $ differs from $ \ket{\chi_f} $ only in the occupation number $ n_i $ (that has polarization $ p_i $) reduced by one. This is in contrast to $ T_{fs}^{(2)} $, where all terms will demand a difference between  $ \ket{\chi_s} $ and $ \ket{\chi_f} $ in exactly two phonon modes. Therefore, in Eq.~\ref{equ:Fermi} no states $ \ket{s} $ and $ \ket{f} $ exist that would give a non-vanishing $ T_{fs}^{(1)} $ and $ T_{fs}^{(2)} $ at the same time. This allows us to separate the overall hopping rate into a one- and a two-phonon contribution as
\begin{equation}\label{equ:khoppSperation}
	k_{\uparrow} = k_{\uparrow,1} + k_{\uparrow,2} + ...\,.
\end{equation}
We then find for
\begin{align}
	k_{\uparrow,1} &= \frac{2\pi}{\hbar} \sum_{f}  \abs{T_{fs}^{(1)}}^2 \delta(\hbar\Delta_\perp+\epsilon_f-\epsilon_s) \\ \nonumber
	&= 2\pi\hbar \sum_{f} \Bigg| \sum_{i} \begin{aligned}[t] &\delta_{p_i,\y}\lambda_i \Big[
		\sqrt{n_{s,i}}\delta_{f,si-}\delta(\hbar\Delta_\perp-\hbar\omega_i)\\ \nonumber
		&+\sqrt{n_{s,i}+1}\delta_{f,si+}\delta(\hbar\Delta_\perp+\hbar\omega_i)\Big] \Bigg| ^2\,,
	\end{aligned}
\end{align}
where the second term vanishes since $ \Delta_\perp,\omega_i >0 $.
Note that for each $ f $ there is at most one term in the sum over $ i $ that is non-zero. On the other hand, for each $ i $ there is exactly one $ f $. Thus, for any $ F_i $
\begin{equation}\label{equ:sumCollapse}
	\sum_{f} \Bigg| \sum_{i} \delta_{f,si+/-} F_i \Bigg| ^2 = \sum_{i} \abs{F_i}^2\,.
\end{equation}
Instead of summing over all modes $ i $ we integrate over all energies $ \epsilon $ and use the average $ \overline{\abs{F_i}^2} $ of all modes $ i $ with energy $ \hbar \omega_i = \epsilon $.
Inserting the identity $ \int_{0}^{\Omega} \delta(\epsilon-\hbar \omega_i) d\epsilon $, with the cutoff energy for E-symmetric phonons $ \Omega $ (see discussion in Sec.~\ref{sec:probing}), we find
\begin{align}\label{equ:k1hoppupWithSpectralDensity}
	%\begin{aligned}
        \nonumber
		k_{\uparrow,1} &= \begin{aligned}[t] 2\pi\hbar \int_{0}^{\Omega} \sum_{i} & \delta_{p_i,\y}\abs{\lambda_i}^2\delta(\epsilon-\hbar \omega_i) n_{s,i} \\ \nonumber
			&\cdot \delta(\hbar\Delta_\perp-\hbar\omega_i) d\epsilon \end{aligned}\\ \nonumber
		&= 2\pi\hbar \frac{2}{\pi\hbar} \int_{0}^{\Omega} J_y(\epsilon) n(\epsilon) \delta(\hbar\Delta_\perp-\epsilon) d\epsilon \\
		&= 4 J_y(\hbar\Delta_\perp) n(\hbar\Delta_\perp)\,,
	%\end{aligned}
\end{align}
where the Bose–Einstein distribution
\begin{equation}\label{equ:nBose}
	n(\epsilon) = \frac{1}{e^{\epsilon/\kB T}-1}
\end{equation}
describes the thermal occupation of phonon modes with energy $ \epsilon $ at temperature $ T $ in $ \ket{\chi_s} $.
Here, the polarization-specific phonon spectral density of E-symmetric phonons
\begin{align}\label{equ:JDefinition}
	%\begin{aligned}
        \nonumber
		J_{\x/\y}(\epsilon) &= \frac{\pi\hbar}{2} \sum_{i} \delta_{p_i,\x/\y}\abs{\lambda_i}^2\delta(\epsilon-\hbar \omega_i) \\
		&= \frac{\pi\hbar}{2} \rho(\epsilon) \left.\overline{\abs{\lambda_{i}}^2}\right|_{p_i=\x/\y,\hbar\omega_i=\epsilon}
	%\end{aligned}
\end{align}
was introduced with the phonon modes density $ \rho(\epsilon) $.
Assuming a linear dispersion relation and a wavelength of the acoustic phonons much larger than the lattice spacing, the Debye model gives $ \rho(\epsilon) \propto \epsilon^2 $ and the coupling strength can be approximated~\cite{abtew11, goldman15prb} to be $ \lambda_i \propto \sqrt{\epsilon} $.
Then, the spectral density is independent of polarization and given by
\begin{align}\label{equ:JlinearApprox}
	J(\epsilon) = \eta \epsilon^3\,.
\end{align}
Here, $ \eta $ is a measure of the coupling strength to phonons.
Finally, we find for the one-phonon orbital hopping rate $ \Ey \rightarrow \Ex $:
\begin{align}\label{equ:k1hoppUp}
	%\begin{aligned}
		k_{\uparrow,1}(T, \dperp) &= 4 \eta \left[\hbar\Delta_\perp\right]^3 n(\hbar\Delta_\perp) \\ \nonumber
		&\approx 32 \eta h^3 \dperp^3  n(2 \dperp h)
	%\end{aligned}
\end{align}
The rates for the opposite direction, $ \Ex \rightarrow \Ey $, can be found from the detailed balance ratio
$k_{\uparrow,1/2}/k_{\downarrow,1/2} = \exp[-\hbar\Delta_\perp/(\kB T)]$ (Eq.~\ref{equ:detailedBallance})
which, as a consequence of Eq.~\ref{equ:khoppSperation}, holds for both the one-phonon and two-phonon processes separately. The result was given in Eq.~\ref{equ:k1hoppDown}:
\begin{equation*}%\label{equ:k1hoppDown}
		k_{\downarrow,1}(T, \dperp) \approx 32 \eta h^3 \dperp^3 \left[ n(2 \dperp h) + 1\right]
\end{equation*}
For elevated temperatures $ \kB T \gg 2 \dperp h $ ($ \SI{5}{\kelvin} \:\widehat{=}\: \SI{50}{\giga\hertz} $ strain), the up- and down-rates of equations \ref{equ:k1hoppUp} and \ref{equ:k1hoppDown} become the same and are, to second order in $ 2 \dperp h / \kB T $, given by~\cite{goldman15prb}:
\begin{equation}\label{equ:k1hopp_TinfLimit}
	k_{1} (T \gg 2 \dperp h/\kB, \dperp) \approx 16 \eta h^2 \dperp^2 \kB T
\end{equation}
%\vspace{-1cm}

\begin{widetext}
\subsection{Two phonon processes}
\label{app:2phon}
Next, we derive the second order transition matrix element in Eq.~\ref{equ:Tfs2}, again for going from the lower to the upper branch, as given in Eq.~\ref{equ:start} and \ref{equ:end}:
% \begin{alignat}{4}
% 	T_{fs}^{(2)} = \sum_{m} & \frac{\hbar^2}{E_s-E_m} \\ \nonumber
%             &\cdot \Bigg( \sum_{i} \Big[
% 					&&\delta_{p_i,\x}\lambda_i\delta_{O_m,\x}\mel{\chi_f}{a_i+a_i^\dagger}{\chi_m} \\ \nonumber 
% 					&&&-\delta_{p_i,\y}\lambda_i\delta_{O_m,\y}\mel{\chi_f}{a_i+a_i^\dagger}{\chi_m} \Big]\Bigg)\\ \nonumber 
%             &\cdot \Bigg( \sum_{j} \Big[
% 					&&-\delta_{p_j,\x}\lambda_j\delta_{O_m,\y}\mel{\chi_m}{a_j+a_j^\dagger}{\chi_s} \\ \nonumber 
% 					&&&-\delta_{p_j,\y}\lambda_j\delta_{O_m,\x}\mel{\chi_m}{a_j+a_j^\dagger}{\chi_s} \Big]\Bigg)
% \end{alignat}
Using the same relations as before (Eq.~\ref{equ:Tfs1_a_adagg}), this yields:
%!%\begin{widetext}
\begin{align}
	T_{fs}^{(2)}
				&= \sum_{m}\frac{\hbar^2}{E_s-E_m} \begin{aligned}[t] &\Bigg( \sum_{i} \Big[ 
					\delta_{p_i,\x}\lambda_i\delta_{O_m,\x}\mel{\chi_f}{a_i+a_i^\dagger}{\chi_m}
					-\delta_{p_i,\y}\lambda_i\delta_{O_m,\y}\mel{\chi_f}{a_i+a_i^\dagger}{\chi_m} \Big]\Bigg) \nonumber \\
					\cdot & \Bigg( \sum_{j} \Big[
					-\delta_{p_j,\x}\lambda_j\delta_{O_m,\y}\mel{\chi_m}{a_j+a_j^\dagger}{\chi_s}
					-\delta_{p_j,\y}\lambda_j\delta_{O_m,\x}\mel{\chi_m}{a_j+a_j^\dagger}{\chi_s} \Big]\Bigg)
				\end{aligned} \\
				&= \sum_{m} \hbar \begin{aligned}[t] &\Bigg( \sum_{i} \lambda_i \Big[ \begin{aligned}[t]
						& \underbrace{\sqrt{n_{f,i}+1}\delta_{m,fi+}\delta_{p_i,\x}\delta_{O_m,\x}}_{\tcircle{~1~}}
						 \underbrace{-\sqrt{n_{f,i}+1}\delta_{m,fi+}\delta_{p_i,\y}\delta_{O_m,\y}}_{\tcircle{~2~}}\\
						& \underbrace{-\sqrt{n_{f,i}}\delta_{m,fi-}\delta_{p_i,\x}\delta_{O_m,\x}}_{\tcircle{~3~}}
						 \underbrace{-\sqrt{n_{f,i}}\delta_{m,fi-}\delta_{p_i,\y}\delta_{O_m,\y}}_{\tcircle{~4~}} \Big]\Bigg)
					\end{aligned} \\
					&\cdot\Bigg( \sum_{j} \lambda_j \Big[ \begin{aligned}[t]
						& \underbrace{-\sqrt{n_{s,j}}\delta_{m,sj-}\delta_{p_j,\x}\delta_{O_m,\y} \frac{1}{\omega_j}}_{\tcircle{~a~}}
						 \underbrace{-\sqrt{n_{s,j}}\delta_{m,sj-}\delta_{p_j,\y}\delta_{O_m,\x} \frac{1}{-\Delta_\perp+\omega_j}}_{\tcircle{~b~}}\\
						& \underbrace{-\sqrt{n_{s,j}+1}\delta_{m,sj+}\delta_{p_j,\x}\delta_{O_m,\y} \frac{1}{-\omega_j}}_{\tcircle{~c~}}
						 \underbrace{-\sqrt{n_{s,j}+1}\delta_{m,sj+}\delta_{p_j,\y}\delta_{O_m,\x} \frac{1}{-\Delta_\perp-\omega_j}}_{\tcircle{~d~}} \Big]\Bigg)
					\end{aligned}
				\end{aligned} \\ \nonumber
\end{align}
%!%\end{widetext}
Since both factors specify attributes of the intermediate state $ m $, in the factored-out equation less than 16 terms remain. For example \textcircled{\scriptsize 1}\textcircled{\scriptsize  a}: $ \delta_{O_m,\x}\delta_{O_m,\y} = 0 $.
On the remaining terms, we will use relations like
\begin{align}
    \delta_{m,fi+}\delta_{m,sj-} = \delta_{fi+,sj-} \delta_{m,sj-} = \delta_{f,si-j-} \delta_{m,sj-}
\end{align}
% \[
% \delta_{m,fi+}\delta_{m,sj-} = \delta_{fi+,sj-} \delta_{m,sj-} = \delta_{f,si-j-} \delta_{m,sj-}
% \]
and for that specific example
\begin{align}
    n_{f,i} = n_{s,i} - 1\,.
\end{align}
% \[
% n_{f,i} = n_{s,i} - 1\,.
% \]
From the above, one can see that the one intermediate state $ m $ for which each term is non-zero is already specified by $ j $ and the $ \delta_{O_m,\x/\y} $. To that end, the sum over $ m $ can simply be dropped:

%!%\begin{widetext}
\begin{align}%\label{equ:T2fs}
	%\begin{aligned}
		T_{fs}^{(2)} = \sum_{i,j} \hbar \lambda_i \lambda_j \Big[
		\sqrt{n_{s,i}}\sqrt{n_{s,j}}\delta_{p_i,\x}\delta_{p_j,\y}\frac{-1}{-\Delta_\perp+\omega_j}&\delta_{f,si-j-} &\tcircle{1}\tcircle{b}\\ \nonumber
		+\sqrt{n_{s,i}}\sqrt{n_{s,j}+1}\delta_{p_i,\x}\delta_{p_j,\y}\frac{-1}{-\Delta_\perp-\omega_j}&\delta_{f,si-j+} &\tcircle{\scriptsize 1}\tcircle{\scriptsize d}\\ \nonumber
		+\sqrt{n_{s,i}}\sqrt{n_{s,j}}\delta_{p_i,\y}\delta_{p_j,\x}\frac{1}{\omega_j}&\delta_{f,si-j-} &\tcircle{\scriptsize 2}\tcircle{\scriptsize a}\\ \nonumber
		+\sqrt{n_{s,i}}\sqrt{n_{s,j}+1}\delta_{p_i,\y}\delta_{p_j,\x}\frac{1}{-\omega_j}&\delta_{f,si-j+} &\tcircle{\scriptsize 2}\tcircle{\scriptsize c}\\ \nonumber
		+\sqrt{n_{s,i}+1}\sqrt{n_{s,j}}\delta_{p_i,\x}\delta_{p_j,\y}\frac{-1}{-\Delta_\perp+\omega_j}&\delta_{f,si+j-} &\tcircle{\scriptsize 3}\tcircle{\scriptsize b}\\ \nonumber
		+\sqrt{n_{s,i}+1}\sqrt{n_{s,j}+1}\delta_{p_i,\x}\delta_{p_j,\y}\frac{-1}{-\Delta_\perp-\omega_j}&\delta_{f,si+j+} &\tcircle{\scriptsize 3}\tcircle{\scriptsize d}\\ \nonumber
		+\sqrt{n_{s,i}+1}\sqrt{n_{s,j}}\delta_{p_i,\y}\delta_{p_j,\x}\frac{1}{\omega_j}&\delta_{f,si+j-} &\tcircle{\scriptsize 4}\tcircle{\scriptsize a}\\ \nonumber
		+\sqrt{n_{s,i}+1}\sqrt{n_{s,j}+1}\delta_{p_i,\y}\delta_{p_j,\x}\frac{1}{-\omega_j}&\delta_{f,si+j+} \Big] &\tcircle{\scriptsize 4}\tcircle{\scriptsize c}
	%\end{aligned}
\end{align}
\end{widetext}

There are three kinds of processes:

(i) Terms \textcircled{\scriptsize  1}\textcircled{\scriptsize  d}, \textcircled{\scriptsize  2}\textcircled{\scriptsize  c}, \textcircled{\scriptsize  3}\textcircled{\scriptsize  b}, \textcircled{\scriptsize  4}\textcircled{\scriptsize  a} describe a two-phonon Raman process in which one phonon is absorbed and one is emitted.
 
(ii) Terms \textcircled{\scriptsize  3}\textcircled{\scriptsize  d}, \textcircled{\scriptsize  4}\textcircled{\scriptsize  c} are processes in which two phonons are emitted. Since the orbital state needs energy $ \hbar \Delta_\perp $ for the transition $ \Ey \rightarrow \Ex $ in consideration here, these processes cannot be energy conserving and thus do not contribute.
 
(iii) Terms \textcircled{\scriptsize  1}\textcircled{\scriptsize  b}, \textcircled{\scriptsize  2}\textcircled{\scriptsize  a} are processes in which two phonons are absorbed. Since their sum of energy has to be $ \hbar \Delta_\perp $, there are only few modes that contribute.

To determine the significance of processes where two phonons are absorbed ($ \Ey \rightarrow \Ex $) or two are emitted ($ \Ex \rightarrow \Ey $), we compare their rates with the respective Raman process.
For most strains studied here ($ \dperp < \SI{100}{\giga\hertz} $), and for all temperatures where the two-phonon process is relevant ($ T > \SI{10}{\kelvin} $), the contribution to the hopping rate is below $ \SI{1}{\percent} $. We will therefore generally neglect two phonon emission and absorption processes.

%We compare the rates of the emission of two phonons with the respective Raman process. This rate (terms \textcircled{\scriptsize  3}\textcircled{\scriptsize  d}, \textcircled{\scriptsize  4}\textcircled{\scriptsize  c}) is stronger but similar to the absorption of two phonons and arises when going from $\Ex$ to $\Ey$ instead. % (Eq.~\ref{equ:k2hoppUp} and \ref{equ:detailedBallance}). 
%For most strains studied here ($ \dperp < \SI{100}{\giga\hertz} $), and for all temperatures where the two-phonon process is relevant ($ T > \SI{10}{\kelvin} $), the contribution to the hopping rate is below $ \SI{1}{\percent} $. We will therefore neglect two phonon emission and absorption processes.

We now determine the orbital hopping rate for the two-phonon Raman process. In terms \textcircled{\scriptsize  1}\textcircled{\scriptsize  d} and \textcircled{\scriptsize  2}\textcircled{\scriptsize  c} one can swap the names of indices $ i $ and $ j $ to make their appearance similar to terms \textcircled{\scriptsize  3}\textcircled{\scriptsize b} and \textcircled{\scriptsize 4}\textcircled{\scriptsize a}. Then, $ i $ is the emitted phonon (``$ i+ $'') and $ j $ the absorbed (``$ j- $''):
\begin{align}
	%\begin{aligned}
		\tilde{T}_{fs}^{(2)} = \sum_{i,j} & \hbar \lambda_i\lambda_j \sqrt{n_{s,i}+1}\sqrt{n_{s,j}} \delta_{f,si+j-} \\ \nonumber
		&\cdot \Big[ \begin{aligned}[t]
			&\delta_{p_i,\y}\delta_{p_j,\x}\frac{-1}{-\Delta_\perp-\omega_i} &\tcircle{~1~}\tcircle{~d~}\\ \nonumber
			&+\delta_{p_i,\x}\delta_{p_j,\y}\frac{1}{-\omega_i} &\tcircle{~2~}\tcircle{~c~}\\
			&+\delta_{p_i,\x}\delta_{p_j,\y}\frac{-1}{-\Delta_\perp+\omega_j} &\tcircle{~3~}\tcircle{~b~}\\ \nonumber
			&+\delta_{p_i,\y}\delta_{p_j,\x}\frac{1}{\omega_j} \Big] &\tcircle{~4~}\tcircle{a}
		\end{aligned}
	%\end{aligned}
\end{align}
Fermi's Golden Rule requires the conservation of energy, which means for the process we look at
\begin{equation}
    \Delta_\perp = \omega_j - \omega_i > 0\,.
\end{equation}
% \[ 
% \Delta_\perp = \omega_j - \omega_i > 0\,.
% \]
With that, terms \textcircled{\scriptsize 1}\textcircled{\scriptsize d}/\textcircled{\scriptsize 2}\textcircled{\scriptsize c} and \textcircled{\scriptsize 4}\textcircled{\scriptsize a}/\textcircled{\scriptsize 3}\textcircled{\scriptsize b} will become the same, giving a factor $ 2 $ and two terms in $ \tilde{T}_{fs}^{(2)} $ of which only one can be non-zero for given $i,j$ due to the required polarization.
Relation \ref{equ:sumCollapse} also holds here for $ \sum_{i,j} $ and we use the spectral density \ref{equ:JDefinition} and \ref{equ:JlinearApprox} twice with $ \delta(\epsilon_1-\hbar\omega_i) $ and $ \delta(\epsilon_2-\hbar\omega_j) $. This gives:
\begin{align}
	k_{\uparrow,2} &= \frac{2\pi}{\hbar} \sum_{f}  \abs{\tilde{T}_{fs}^{(2)}}^2 \delta(\hbar\Delta_\perp+\epsilon_f-\epsilon_s) \nonumber  \\
	&= \begin{aligned}[t] \frac{2\pi}{\hbar} \left(\frac{2}{\pi\hbar}\right)^2 4\hbar^2 \int_{0}^{\Omega}\int_{0}^{\Omega}
		& J(\epsilon_1) J(\epsilon_2) \nonumber  \\
		& \cdot \left[n(\epsilon_1)+1\right]n(\epsilon_2) \\
		& \cdot \left( \abs{\frac{\hbar^2}{\hbar\Delta_\perp+\epsilon_1}}^2 + \abs{\frac{\hbar^2}{\epsilon_1}}^2 \right) \\
		& \cdot \delta(\hbar\Delta_\perp+\epsilon_1-\epsilon_2) d\epsilon_1 d\epsilon_2
	\end{aligned}\\
	&= \begin{aligned}[t] \frac{32\hbar}{\pi}  \int_{\hbar\Delta_\perp}^{\Omega}
		& J(\epsilon_2-\hbar\Delta_\perp) J(\epsilon_2) \\
		& \cdot \left[n(\epsilon_2-\hbar\Delta_\perp)+1\right]n(\epsilon_2) \\
		& \cdot \left( \abs{\frac{1}{\epsilon_2}}^2 + \abs{\frac{1}{\epsilon_2-\hbar\Delta_\perp}}^2 \right) d\epsilon_2
	\end{aligned}
\end{align}
Likewise, with the detailed balance \ref{equ:detailedBallance} we find for the rate $ \Ex \rightarrow \Ey $:
\begin{align}
		k_{\downarrow,2} = \frac{32\hbar}{\pi}  \int_{\hbar\Delta_\perp}^{\Omega} 
		& J(\epsilon-\hbar\Delta_\perp) J(\epsilon) \left[n(\epsilon)+1\right]n(\epsilon-\hbar\Delta_\perp) \nonumber \\
		& \cdot \left( \frac{1}{\epsilon^2} + \frac{1}{(\epsilon-\hbar\Delta_\perp)^2} \right) d\epsilon
\end{align}
Using the expressions for $ J(\epsilon) $ (\ref{equ:JlinearApprox}) and $ n(\epsilon) $ (\ref{equ:nBose}) and the substitutions $ x = \epsilon/(\kB T) $ and $ x_\perp = \hbar\Delta_\perp/(\kB T) $ (\ref{equ:splitting}) we find, for example, $ k_{\uparrow,2}(T,\dperp) $ as given in Eq.~\ref{equ:k2T5} and \ref{equ:I}:
\begin{align*}
	%\begin{aligned}
		k_{\uparrow,2}(T,\dperp) &= \frac{64\hbar}{\pi} \eta^2 \kB^5 T^5 I(T, \dperp)\\
        I(T, \dperp) &= \int_{x_\perp}^{\Omega/\kB T} \frac{e^{x}x(x-x_\perp)\left[x^2+(x-x_\perp)^2\right]}{2 \left(e^{x}-1\right) \left(e^{x-x_\perp}-1\right)} dx
	%\end{aligned}
\end{align*}
Using the detailed balance Eq.~\ref{equ:detailedBallance}, one can directly obtain $ k_{\downarrow,2}(T,\dperp) $, which is similar to $ k_{\uparrow,2}(T,\dperp) $ at elevated temperature $ \kB T \gg 2 \dperp h $.
In Ref.~\cite{ernst2023}, we plot the hopping rates derived in this section as a function of temperature for various strain values.

The hopping processes discussed in this section are like a $ T_1 $ process on the orbital state towards the mixed state in the detailed balance equilibrium. It will therefore inherently also cause a $ T_2 $ process on the coherent orbital state, which has the same rate as $ k_{\downarrow/\uparrow} $.
Coupling to $\text{A}_1$-symmetric phonon modes also contributes to the overall orbital dephasing~\cite{plakhotnik15}, which we did not include in our model.
These modes can only influence the PL in case of resonant optical excitation into levels of the \textit{zf basis}, i.e. in very low strain conditions, as \textit{zf basis} states are coherent superpositions of the orbital eigenstates (Eq.~\ref{equ:TzfSz}). This might slightly influence the picture given in Fig.~\ref{fig:ISC}.

\section{Jump Operators}
\label{app:jump}
As an example for incoherent transition rates, there are six jump operators to describe the optical decay. An example expressed in \textit{hf basis} (Eq.~\ref{equ:oe_state}) is
\begin{equation}\label{equ:Lkr}
	\hat{L}_{r,+1} = \sqrt{k_r} \ket{+1}\bra{E_{\x,+1}}\,.
\end{equation}
On the other hand, for example, there is one jump operator to describe the ISC of state $ \ket{A_1} $. Thus, expressed in \textit{zf basis} (Eq.~\ref{equ:zf_state})
\begin{equation}\label{equ:LA1}
\hat{L}_{\text{ISC},A_1} = \sqrt{k_{A_1}} \ket{ss}\bra{A_1}\,.
\end{equation}
To use  $ \hat{L}_{r,+1} $ and $ \hat{L}_{\text{ISC},A_1} $ in the master Eq.~\ref{equ:ME}, the matrix representation of it in \textit{hf basis} or \textit{zf basis}, respectively, needs to be basis transformed to the \textit{$e_z$ basis} (basis in which $H$ is written) via the transformation matrices given in equations \ref{equ:TzfSz} and \ref{equ:ToeSz}.
We note that these jump operators destroy coherences between states. For example, an initial state $ \ket{\psi} = 1/\sqrt{2} \left( \ket{E_{\x,+1}} + \ket{E_{\x,-1}} \right) $ will decay to a 50:50 classical mixture of $ \ket{+1} $ and $ \ket{-1} $ under the influence of jump operators $ \hat{L}_{r,+1} $ and $ \hat{L}_{r,-1} $. This behavior is intended for all rates in this model with the exception of the orbital branch hopping $ k_{\downarrow/\uparrow} $.
Here, a jump only leads to a decay of the orbital state and therefore to the destruction of coherences in the orbital subspace $ \mathcal{H}_{\text{orbit}} $.
Crucially, coherences in the spin subspace $ \mathcal{H}_{\text{spin}} $ are maintained when jumping from one orbital branch to the other.
To that end, there are two jump operators for orbital branch hopping:
\begin{align}
	\hat{L}_{\downarrow}^\text{ES} &= \sqrt{k_{\downarrow}} \op{\y}{\x} \otimes \hat{\mathbb{I}}_3 \label{equ:LHoppx2y}\\
	\hat{L}_{\uparrow}^\text{ES} &= \sqrt{k_{\uparrow}} \op{\x}{\y} \otimes \hat{\mathbb{I}}_3 \label{equ:LHoppy2x}
\end{align}
To arrive at the representation in the full \textit{hf basis}, the matrix representation has to be extended by $ 0 $-block matrices.

We note that while maintaining spin-state coherences during orbital hopping is crucial, adding ES spin-state decoherence to our model is found to not alter the temperature-dependent behavior noticeably. A decoherence of around $ \SI{10}{\nano\second} $ was observed at~\cite{fuchs10} and above~\cite{plakhotnik15} room temperature and attributed to the motional narrowing also caused by the orbital hopping process.
We therefore assume no spin $ T_2 $ process in this study.

In this work, we assume off-resonant optical excitation but with respective adaptions in the jump operators, also resonant laser excitation into selected eigenstates of the excited state can be simulated, as done for Fig.~\ref{fig:ISC}.

\section{Comparison with Goldman \etal}
\label{sec:SIcomparisonISC}
We look at the ISC rates and their temperature dependence at low temperature to verify that our model yields the behavior observed in previous work.
To obtain the ISC rates, we mimic the experiments by \citet{goldman15}. First, we initialize the system in the respective eigenstates. Then, we evolve the state in time and calculate the PL from Eq.~\ref{equ:PL}. Finally, we fit a single exponential decay to it with a decay rate of $ k_r $ (known) plus the effective ISC rate $k_{\text{ISC}}$ (wanted). We use parameters as found by Goldman \etal, in particular $ \eta = \SI{276.5}{\per\micro\second\per\cubic\milli\electronvolt}$, but we do not use their $ T_0 = \SI{4.4}{\kelvin} $ shift.
We note that fitting with a single-exponential decay is systematically not correct, as around $ \SI{15}{\kelvin} $ (where averaging of the ES levels ISC rates happens as the optical lifetime is similar to the orbital hopping rate) contributions form the different states can be observed. The ISC rates plotted here should thus be seen as an approximation.
The ISC rate difference of $ E_1 $ and $ E_2 $ at low temperature originates from a finite strain splitting of $ 2 \dperp = \SI{3.9}{\giga\hertz} $. 
This behavior could not be resolved in the experiment since the states are spectrally too close together for explicit resonant excitation~\cite{goldman15}.
Also, we note that contributions from $A_1$-symmetric phonon modes might matter here, which are not considered in this paper (see discussion at the end of Appendix~\ref{app:2phon}).

\section{Comparison with Plakhotnik \etal}
\label{sec:SItrf}
We employ our model at room temperature and very high strain to examine the predicted behavior.
\citet{plakhotnik14} derived that in the absence of a magnetic field, an additional term
\begin{alignat}{4}\label{equ:Htrf}
	%\begin{aligned}
		\hat{H}_{\text{TRF}}/h = & - D^\perp_{\text{ES}} \mathcal{R}(T, \dperp) %\begin{aligned}[t] 
        \Big[ 
			&&\cos{\phi_\delta} \left(\hat{S}_y^2 - \hat{S}_x^2 \right) \\ \nonumber
			&&&+ \sin{\phi_\delta} \left(\hat{S}_y \hat{S}_x + \hat{S}_x \hat{S}_y \right)\Big]
		%\end{aligned}
        \\ \nonumber
		&- \lambda^\perp_{\text{ES}} \mathcal{R}(T, \dperp) %\begin{aligned}[t] 
        \Big[ 
			&&\cos{\phi_\delta} \left(\hat{S}_x \hat{S}_z + \hat{S}_z \hat{S}_x \right) \\ \nonumber
			&&&+ \sin{\phi_\delta} \left(\hat{S}_y \hat{S}_z + \hat{S}_z \hat{S}_y \right)\Big]
		%\end{aligned}
	%\end{aligned}
\end{alignat}
with the temperature reduction factor
\begin{equation}\label{equ:trf}
	\mathcal{R}(T, \dperp) = \frac{e^{\hbar \Delta_\perp/\kB T}-1}{e^{\hbar \Delta_\perp/\kB T}+1}
\end{equation}
and the strain splitting $\hbar \Delta_\perp \approx 2 h \dperp$ from Eq.~\ref{equ:splitting} has to be added to the Hamiltonian of the ES after orbital averaging (Eq.~\ref{equ:HspinSubSpace}). We corrected a sign error in Eq.~\ref{equ:Htrf} compared to the derivation given in Ref.~\cite{plakhotnik14}.
In this, we still assume that constants as $ D^\perp_{\text{ES}} $ (see Tab.~\ref{tab:1}) have no temperature and strain dependence. Such intrinsic dependencies are discussed in Ref.~\cite{plakhotnik14} but ignored here, as they affect all models equally and thus do not matter for the comparison done here.

We note that in the presence of a magnetic field, Eq.~\ref{equ:Htrf} is not analytically correct due to the orbital $g_l$-factor term in the ES Hamiltonian (Eq.~\ref{equ:Hes}). But the ES LAC at $B_z \approx \SI{51}{\milli\tesla}$ shows a very small shift of $\Delta B_{\text{LAC}} \approx \SI{2}{\micro\tesla}$ compared to Eq.~\ref{equ:Htrf} at a strain of $\dperp = \SI{1}{\tera\hertz}$ and at $\SI{300}{\kelvin}$ with $ g_l = 0.1 $~\cite{rogers09} ($\Delta B_{\text{LAC}} \approx \SI{20}{\micro\tesla}$ with $g_l = 1.0 $). Therefore, Eq.~\ref{equ:Htrf} still constitutes a good approximation as long as the $g_l B_z$ term is small.
But at high magnetic fields and if a significant strain dependence of the orbital $g_l$-factor exists, as recent measurements by \citet{happacher22} indicate, the $g_l B_z$ term could become relevant.
In general, the ES averaged Hamiltonian reads~\cite{plakhotnik14}
\begin{align}\label{equ:Htrf_correct}
    %\begin{aligned}
        \nonumber
		H_{ES,\,\text{avg}} = &H_{\text{spin}} + H_{\text{TRF}}\\
		= &D_{\text{ES}}^\parallel \left( S_z^2 - \tfrac{2}{3} \mathbb{I}_3 \right)
		-\lambda_{\text{ES}}^\parallel \sigma_2 S_z \\ \nonumber
		&+D_{\text{ES}}^\perp \left[ \sigma_3 \left( S_y^2 -S_x^2 \right) - \sigma_1 \left( S_y S_x + S_x S_y \right)  \right] \\ \nonumber
		&+\lambda_{\text{ES}}^\perp \left[ \sigma_3 \left( S_x S_z + S_z S_x \right) - \sigma_1 \left( S_y S_z + S_z S_y \right)  \right] \\ \nonumber
		&+\muB g_{\text{ES}} \vec{S} \cdot \vec{B}
  		+ \muB g_{l} B_z \sigma_2 \mathbb{I}_3 \\ \nonumber
        &+ d_{\text{ES}}^\perp \xi_x \sigma_3 \mathbb{I}_3 - d_{\text{ES}}^\perp \xi_y \sigma_1 \mathbb{I}_3 + d_{\text{ES}}^\parallel \xi_z \mathbb{I}_3
    %\end{aligned}
\end{align}
with $\sigma_{1/2/3} = \Tr\left( \sigma_{x/y/z} \rho_{\text{orbit}} \right)$,
\begin{align}
	\rho_{\text{orbit}} =  & T^{\text{orbit}}_{\textit{hf} \rightarrow e_z}
		\mqty(
		e^{-\hbar \Delta_\perp/\kB T} &    0 \\
		0 &    1 \\
		) \cdot {T^{\text{orbit}}_{\textit{hf} \rightarrow e_z}}^{-1} \\ \nonumber
        & \cdot \left(1+e^{-\hbar \Delta_\perp/\kB T}\right)^{-1}
\end{align}
and $ T^{\text{orbit}}_{\textit{hf} \rightarrow e_z} $ as used in Eq.~\ref{equ:ToeSz}. All $\mathbb{I}_3$ terms in Eq.~\ref{equ:Htrf_correct} cause an overall energy shift of the ES and are thus irrelevant to our analysis.

\section{Implementation of the Model}
\label{sec:impl}
Our model is openly accessible via github~\cite{ernst_nvratemodel_2023}, is written in Python, and was optimized for fast computation speed by the open access numba package \cite{numba}. It has $ 10 $ levels (c.f. Eq.~\ref{equ:Sz_state}). Thus, the FLS has a dimension of $ 100 $ with $ \tilde{\mathcal{L}} \in \mathbb{C}^{100\cross100} $. It is computationally expensive to calculate the matrix exponential $ \exp\big(\tilde{\mathcal{L}}t\big) $ in Eq.~\ref{equ:rho} for each time step in a $ \text{PL}(t) $ time trace of Eq.~\ref{equ:PL}. Therefore, we first calculate $	U_{\Delta t} = \exp \big(\tilde{\mathcal{L}} \Delta t\big) $ for a single time step $ \Delta t $. We then obtain the time trace $ \text{PL}_n $ over $ t_n=n \Delta t $, $ n \in [0, N] $, from $ \vec{\rho}_n = U_{\Delta t} \vec{\rho}_{n-1} $ at reduced computational cost. The accumulated numerical errors were found to have a negligible effect on our results.

\input{"references.bbl"}
%\bibliography{library}
	
\end{document}

%% file: references.bbl
%apsrev4-2.bst 2019-01-14 (MD) hand-edited version of apsrev4-1.bst
%Control: key (0)
%Control: author (8) initials jnrlst
%Control: editor formatted (1) identically to author
%Control: production of article title (0) allowed
%Control: page (0) single
%Control: year (1) truncated
%Control: production of eprint (0) enabled
%